\documentclass[a4paper,cite]{article}
\usepackage{latexsym}
\newcommand{\lbl}[1]{\label{eq:#1}}
\newcommand{ \rf}[1]{(\ref{eq:#1})}

\newcommand{\be}{\begin{equation}}
\newcommand{\ee}{\end{equation}}
\newcommand{\bea}{\begin{eqnarray}}
\newcommand{\eea}{\end{eqnarray}}
\newcommand{\setl}{\setlength\arraycolsep{2pt}}

\newcommand{\noi}{\noindent}
\newcommand{\nn}{\nonumber}
\newcommand{\ra}{\rightarrow}
\newcommand{\Ra}{\Rightarrow}

\newcommand{\lesssim}{ {\
\lower-1.2pt\vbox{\hbox{\rlap{$<$}\lower5pt\vbox{\hbox{$\sim$}}}}\ }
}
\newcommand{\gtrsim}{ {\
\lower-1.2pt\vbox{\hbox{\rlap{$>$}\lower5pt\vbox{\hbox{$\sim$}}}}\ }
}

\newcommand{\cD}{{\cal D}}

\newcommand{\cH}{{\cal H}}

\newcommand{\cL}{{\cal L}}
\newcommand{\cM}{{\cal M}}

\newcommand{\cO}{{\cal O}}

\newcommand{\cW}{{\cal W}}

\newcommand{\Imm}{\mbox{\rm Im}}
\newcommand{\Ree}{\mbox{\rm Re}}
\newcommand{\Tr}{\mbox{\rm Tr}}
\newcommand{\tr}{\mbox{\rm tr}}

\newcommand{\MeV}{\mbox{\rm MeV}}
\newcommand{\GeV}{\mbox{\rm GeV}}

\newcommand{\with}{\mbox{\rm with}}

\newcommand{\annd}{\mbox{\rm and}}
\newcommand{\foor}{\mbox{\rm for}}

\newcommand{\als}{\alpha_{\mbox{\rm {\scriptsize s}}}}

\newcommand{\muhad}{\mu_{\mbox{\rm {\scriptsize had.}}}}
\newcommand{\GF}{G_{\mbox{\rm {\tiny F}}}}
\newcommand{\MHA}{\mbox{\rm {\tiny MHA}}}

\newcommand{\gL}{\frac{1-\gamma_{5}}{2}}
\newcommand{\gR}{\frac{1+\gamma_{5}}{2}}

\newcommand{\eff}{\mbox{\rm eff}}

\newcommand{\QCD}{\mbox{\rm {\footnotesize QCD}}}

\newcommand{\msb}{\overline{\mbox{\rm\footnotesize MS}}}

\newcommand{\g}{\mbox{\bf g}}

\newcommand{\bQ}{{\bf Q}}

\newcommand{\psls}{\not \! p}
\newcommand{\qsls}{\not \! q}

\newcommand{\stern}{\langle\bar{\psi}\psi\rangle}
\newcommand{\gew}{\mbox{\bf g}_{\mbox{\rm\footnotesize \underline{ew}}}}

\input epsf

\def\theequation{\arabic{section}.\arabic{equation}}

\begin{document}

\begin{titlepage}

\begin{flushright}
%\today
\end{flushright}
\vspace*{1.5cm}
\begin{center}

{\Large \bf $\Delta I=1/2$ and $\epsilon'/\epsilon$ in
Large--$N_c$ QCD}\\[2.0cm]

{\bf Thomas Hambye}$^{a,b}$,
%\footnote{~Present address, ... Pisa},
{\bf Santiago Peris}$^c$ and  {\bf Eduardo de Rafael}$^{a,c,d}$\\[1cm]

$^a$  Centre  de Physique Th{\'e}orique\\
       CNRS-Luminy, Case 907\\
    F-13288 Marseille Cedex 9, France\\[0.5cm]
$^b$ Scuola Normale Superiore\\ 7, piazza dei Cavalieri, 56126 Pisa, Italy.\\
[0.5cm]

$^c$ Grup de F{\'\i}sica Te{\`o}rica and IFAE\\ Universitat
Aut{\`o}noma
de Barcelona, 08193 Barcelona, Spain.\\
[0.5cm]
$^d$ Instituci\'o Catalana de Recerca i Estudis Avan\c{c}ats (ICREA).\\

\end{center}

\vspace*{1.0cm}

\begin{abstract}
We present new results for the matrix elements of
the $Q_6$ and $Q_4$ penguin operators, evaluated in a large--$N_c$
approach which incorporates important
$\cO(N_c^2\frac{n_f}{N_c})$ unfactorized contributions. Our approach
shows analytic matching between short-- and long--distance scale
dependences within dimensional renormalization schemes, such as $\msb$.
Numerically, we find that there is  a large positive  contribution to the
$\Delta I =1/2$ matrix element of $Q_6$ and hence to the direct
CP-violation parameter
$\varepsilon'/\varepsilon$. We also present results for the $\Delta I
= 1/2$ rule in
$K \rightarrow \pi \pi$ amplitudes, which incorporate
the related and important ``eye--diagram'' contributions of
$\cO(N_c^2\frac{1}{N_c})$ from the
$Q_2$ operator (i.e. the penguin--like contraction). The results lead to
an enhancement of the
$\Delta I =1/2$ effective coupling.
The origin of the large unfactorized contributions which we find is
discussed in terms of the relevant scales of the problem.

\end{abstract}

\end{titlepage}

%%%%%%%%%%%%%%%%%%%%%%%%%%%%%%%%%%
\section{\large Introduction}
\lbl{int}

\noi
The so called penguin operators in Particle Physics  have a long history.
Their existence  was first pointed out by the ITEP group~\cite{SVZP75}
who showed that in the process of integrating out heavy fields in the
electroweak theory of the Standard Model, in the presence of QCD
interactions, there appear new local four--quark operators, like e.g.
(summation  over quark colour indices within brackets is understood):
{\setl
\bea\lbl{Q4Q6}
Q_4  =
4\sum_{q=u,d,s}(\bar{s}_{L}\gamma^{\mu}q_{L})
  (\bar{q}_{L}\gamma_{\mu}d_{L})
\quad\  \annd\quad  Q_6  =
-8\sum_{q=u,d,s}(\bar{s}_{L}q_{R})(\bar{q}_{R}d_{L})\,,
\eea}

\noi
besides the two conventional operators
{\setl
\bea\lbl{Q2Q1}
Q_2  =
4(\bar{s}_{L}\gamma^{\mu}u_{L})
  (\bar{u}_{L}\gamma_{\mu}d_{L})
\quad\  \annd\quad  Q_1  =
4(\bar{s}_{L}\gamma^{\mu}d_{L})(\bar{u}_{L}\gamma_{\mu}u_{L})\,,
\eea}

\noi
which had been considered up to then~\cite{GL74,AM74}. These four--quark
operators, modulated by their appropriate Wilson coefficients
$C_{i}(\mu)$, which are functions of the masses of the fields which have
been integrated out and the scale
$\mu$ of whatever renormalization scheme has been used to
carry out this integration, are part of the
effective Hamiltonian which describes the weak interactions of quarks at
intermediate energies of a few GeV~\footnote{ See e.g. the lectures of
A.~Buras in ref.~\cite{Buras97}, where earlier references can also be
found.},
\be\lbl{effH}
\cH_{\eff}(\Delta S=1)=\frac{\GF}{\sqrt{2}}V_{us}^{*}V_{ud}\sum_{i=1}^
{10}C_{i}(\mu)Q_{i}(\mu)\,.
\ee

The ITEP group also claimed that matrix elements of the $Q_6$
penguin operator were particularly important and could be at the
origin of the experimentally well established $\Delta\!I=\!1/2$
enhancement in $K\ra\pi\pi$ transitions. However, the dynamical
mechanism they proposed was shown to violate Current Algebra
Ward identities~\cite{DP84,Oliveretal,Donoghueetal,ChFG86}. In fact,
in the  approximation where the $Q_6$ operator factorizes into a
product of bilinear quark densities, the $K\ra\pi\pi$ matrix
elements turn out to be too small to explain the
$\Delta\!I=\!1/2$ enhancement.

In the meantime, it was also shown that the Wilson coefficient
$C_{6}(\mu)$ acquires a large imaginary part, as a result of the
integration of the heavy $t$ and $b$ quarks and the flavour
mixing structure of the CKM matrix in the Standard Model. As a
result, the evaluation of $K\ra\pi\pi$ matrix elements of the
$Q_6$ operator, besides its potential contribution to $\Delta
I\!=\!1/2$ transitions, has also become of utmost importance,
as an ingredient for a plausible understanding of the
observed size of the CP--violation parameter $\epsilon'/\epsilon$
in the Standard Model.

The effective Lagrangian which describes $\vert\Delta S\vert\!=1$
transitions, like
$K\ra\pi\pi$ and
$K\ra\pi\pi\pi$, in the presence of electromagnetic
interactions and to lowest order in  chiral
perturbation theory ($\chi$PT), $\cO(p^0)$ and
$\cO(p^2)$ in this case, has
the following structure :
\be\lbl{ewcl}
\cL_{\eff}^{\Delta S=1}=-\frac{\GF}{\sqrt{2}}V_{ud}V_{us}^*\left[
\g_{\underline{8}}\cL_{8}+\g_{\underline{27}}\cL_{27}+e^2
\gew\tr\left(U\lambda_
{L}^{(32)}UQ_R
\right)
\right]\,,
\ee
where
\be
\cL_8=\sum_{i=1,2,3}(\cL_{\mu})_{2i}\ (\cL^{\mu})_{i3}\quad\annd\quad
\cL_{27}=\frac{2}{3}(\cL_{\mu})_{21}\ (\cL^{\mu})_{13}+
(\cL_{\mu})_{23}\ (\cL^{\mu})_{11}\,.
\ee
with
\be
\cL_{\mu}=-iF_{0}^2\ U(x)^{\dagger}D_{\mu} U(x)\,,
\ee
and
\be
\lambda_{L}^{(32)}=\delta_{i3}\delta_{j2}\,,\qquad
Q_{L}=Q_{R}=Q=\mbox{\rm diag.}(2/3,-1/3,-1/3)\,.
\ee
The pion decay
coupling constant $F_0$ is the one in the chiral limit where the quark
masses
$u$,
$d$, $s$ are neglected ($F_0\simeq 87~\MeV$).
The matrix field $U$ collects the Goldstone
fields  of the spontaneously broken
chiral symmetry of the QCD Lagrangian with three
massless flavours, and $D_{\mu}U$ denotes the covariant derivative:
$D_{\mu}U\!=\!\partial_{\mu}U-ir_{\mu}U+iUl_{\mu}$, in the
presence of external chiral sources $l_{\mu}$ and $r_{\mu}$ of
left-- and right--handed currents.
Notice that the octet
term proportional to
$\g_{\underline{8}}$ induces pure
$\Delta I=1/2$ transitions, while the term proportional to
$\g_{\underline{27}}$ induces both
$\Delta I=1/2$ and
$\Delta I=3/2$ transitions.
The coupling
constants $\g_{\underline{8}}$, $\g_{\underline{27}}$ and $\gew$ encode
the dynamics of the integrated degrees of freedom in the chiral limit.
These include, not only the heavy quark flavours, but also the hadronic
light flavour states (resonances) other than the Goldstone
fields explicitly present in the $\chi$PT Lagrangian

The coupling
constant $\g_{\underline 8}$ has both a real part  (relevant
for the $\Delta I=1/2$ rule) and an imaginary
part (which induces
direct CP
violation). The main purpose of this work is to present an evaluation of
the contributions to both $\epsilon'/\epsilon$ and $\Delta I=1/2$ induced
by the QCD penguin operators in Eq.~\rf{Q4Q6}, as well as from the so
called {\it eye--like configurations} of the
$Q_2$ operator in Eq.~\rf{Q2Q1}. We shall do this within the framework of
the
$1/N_c$ expansion. The methodology~\footnote{See e.g.
refs.~\cite{PE02,deR02}
for recent reviews.} is the same as the one which has been
applied to other calculations of similar low--energy observables
reported elsewhere~\cite{KPdeR99,PdeR00,KPdeR01,KN02,KPPdeR02,CP03}. It
allows  for important improvements with respect to  earlier calculations
within the framework of the
$1/N_c$ expansion, reported in ref.~\cite{BBG} (at leading order)
and in refs.~\cite{HKPSB98,HKS99,HKPS99,BP99,BP00} (at next--to--leading
order).

The paper is organized as follows.
Section~2 collects the general formula from which the bosonizations
of the penguin operators can be obtained. Section~3 gives the
results for the usual
factorized contributions. Section~4 contains a detailed discussion of the
calculation of the unfactorized contributions of $\cO(N_c n_f)$. Section
5 contains the corresponding analytic results for
the coupling constant $\g_{\underline{8}}$ with both the factorized and
unfactorized contributions. The phenomenological
implications for the $\Delta I = 1/2$ rule and for
$\varepsilon'/\varepsilon$ are presented in sections 6 and 7
respectively. Finally in section 8  we present a discussion of
our results and conclusions.

\vspace{0.7cm}

%%%%%%%%%%%%%%%%%%%%%%%%%%%%%%%%%%%%%%%%%%%%%%%%%%%%%%%%%
%%%%%%%%%%%%%%%%%%%%%%%%%%%%%%%%%%%%%%%%%%%%%%%%%%%%%%%%%

\section{\large Bosonization of the Penguin Operators}
\setcounter{equation}{0}
\lbl{BPO}

\noi
The Dirac operator $\cD_{\chi}$ of the QCD
Lagrangian
\be\lbl{QCDL}
\cL_{\QCD}=-\frac{1}{4}G_{\mu\nu}^{(a)}G^{(a)\mu\nu}+
i\bar{q}\cD_{\chi}q\,,
\ee
in the presence of external chiral sources
$l_{\mu}$, $r_{\mu}$, $\cM$ and
$\cM^{\dagger}$, is defined as follows:

\be\lbl{dirac}
\cD_{\chi} =  \gamma^{\mu}(\partial_{\mu}+iG_{\mu})-
i\gamma^{\mu}\left[l_{\mu}\gL+r_{\mu}\gR\right]
+i\left(\cM\gL+\cM^{\dagger}\gR\right)\,,
\ee

\noi
where $G_{\mu}$ denotes the gluon matrix field, and $G_{\mu\nu}^{(a)}$
the gluon field strength tensor $(a=1,\dots, N_c^2-1)$. The bosonization
of the penguin operators $Q_4$ and $Q_6$, to $\cO(N_c^2)$ and
$\cO(N_c)$,  is then formally defined by the functional
integrals~\cite{PdeR91}

{\setl
\bea\lbl{Q4}
\lefteqn{\langle \bQ_{4}(x)\rangle=4\ \Tr\ \cD_{\chi}^{-1}
(-i)\frac{\delta\cD_{\chi}}{\delta l_{\mu}(x)_{3j}}\Tr\
\cD_{\chi}^{-1} (-i)\frac{\delta\cD_{\chi}}{\delta l^{\mu}(y)_{j2}}}
\nn
\\ & & +4\
\int d^4 y\int\frac{d^4 q}{(2\pi)^4}e^{-iq\cdot (x-y)}\ \Tr
\left(\cD_{\chi}^{-1}(-i)\frac{\delta\cD_{\chi}}{\delta
l_{\mu}(x)_{3j}}\cD_{\chi}^{-1} (-i)\frac{\delta\cD_{\chi}}{\delta
l^{\mu}(y)_{j2}} \right)\,,
\eea}

\noi
and

{\setl
\bea\lbl{Q6}
\lefteqn{\langle \bQ_{6}(x)\rangle=-8\ \Tr\ \cD_{\chi}^{-1}
(-i)\frac{\delta\cD_{\chi}}{\delta \cM^{\dagger}(x)_{3j}}\Tr\
\cD_{\chi}^{-1} (-i)\frac{\delta\cD_{\chi}}{\delta \cM(y)_{j2}}}
\nn
\\ & & +8\
\int d^4 y\int\frac{d^4 q}{(2\pi)^4}e^{-iq\cdot (x-y)}\ \Tr
\left(\cD_{\chi}^{-1}(-i)\frac{\delta\cD_{\chi}}{\delta\cM
{\dagger}(x)_{3j}}\cD_{\chi}^{-1} (-i)\frac{\delta\cD_{\chi}}{\delta
\cM(y)_{j2}} \right)\,,
\eea}

\noi where the trace $\Tr$ here also includes the functional
integration over the {\it planar} gluonic configurations which
leave the quarks at the edge ~\cite{THFT74,W79}. The first terms
in Eqs.~\rf{Q4} and \rf{Q6} correspond to the {\it factorized}
contributions, illustrated in Fig.~1, which are $\cO(N_c^2)$;
while the second terms in Eqs.~\rf{Q4} and \rf{Q6} correspond to
the {\it unfactorized} $\cO(N_c)$ contributions, also illustrated in
Fig.~1. These {\it unfactorized} contributions involve integrals over
the incoming $q$ momenta, which are {\it regularization}
dependent. For consistency, they have to be defined in the same
renormalization scheme as those of the corresponding Wilson coefficients.
With $p$ the conjugate momentum operator, in the absence of the external
chiral sources, the full quark propagator in $x$ space is given by the
expression \be (x\vert
\frac{1}{\cD_{\chi}}\vert y)= (x\vert
\frac{i}{\psls+\gamma^{\alpha}\left[l_{\alpha}\gL+r_{\alpha}\gR\right]
-\cM\gL+\cM^{\dagger}\gR} \vert y) \ee
where
\be\lbl{Gprop} (x\vert
\psls\vert y)=\gamma^{\mu}\left[i\frac{\partial}{\partial x^{\mu}}
-G_{\mu}\right]\delta(x-y)\,. \ee
The bosonization of four--quark
operators in $\chi$PT is then obtained via an appropriate  chiral
expansion in powers of the $l_{\alpha}$, $r_{\alpha}$, $\cM$ and
$\cM^{\dagger}$ external sources in the propagators.

%%%%%%%%%%%% Figure 1 %%%%%%%%%%
\vskip 2pc \centerline{\epsfbox{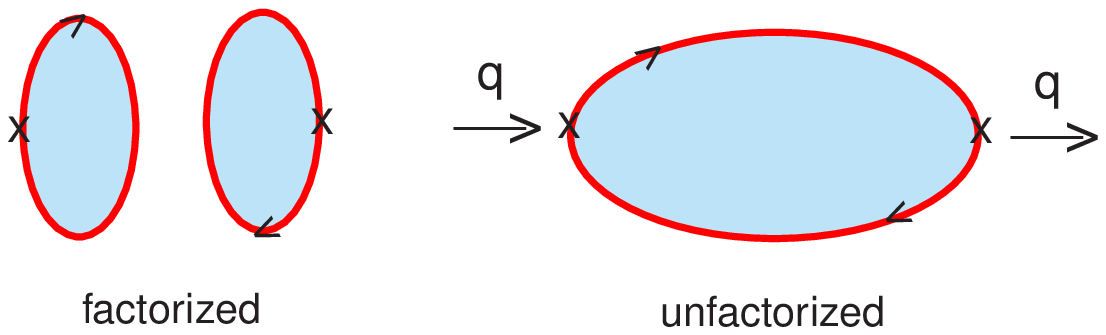}}\vspace*{0.5cm}
%\centerline{
{\bf Fig.~1} {\it Factorized  and unfactorized
contributions corresponding to Eqs.~\rf{Q4} and \rf{Q6}. The crosses
correspond to the explicit variation of the Dirac propagator with
respect to external sources (terms like $\frac{\delta\cD_{\chi}}{\delta
\cM(y)_{j2}}$ in Eq.~\rf{Q6}); the solid lines represent quarks, in the
presence of {\it soft} external chiral sources, propagating in a {\it
planar} gluon background.}
%}
\vskip 2pc
%%%%%%%%%%%%%%%%%%%%%%%%%%%%%%

Let us recall that the spectrum of hadronic states which can contribute
to these Green's functions in QCD in the large--$N_c$
limit~\cite{THFT74,W79} consists of an infinite number of narrow stable
meson states which are flavour nonets. In the real world, however, the
physical hadronic states have widths, which are subleading in $1/N_c$.
The fact that the ratios of widths to masses are of $\cO(1/N_c)$ is at
the very basis of the phenomenological success of the large--$N_c$
approach. Indirectly, this also suggests that in calculating physical
observables, the contributions which stem from narrow states are the
ones which may be potentially important. Notice that in the case of
the bosonization of four--quark operators, the large--$N_c$ QCD
spectrum of narrow states contributes to the {\it factorized}
\underline{as well as} to the {\it unfactorized} patterns. Consequently
one might already expect the {\it unfactorized} piece to be sizeable. Our
calculation shows that this expectation is confirmed and the {\it
unfactorized} contributions from these penguin operators turn out to be
\emph{larger} than those from the {\it factorized} terms. We shall later
argue that this result  becomes rather natural when formulated in terms
of the different hadronic scales involved.

\vspace*{0.3cm}

%%%%%%%%%%%%%%%%%%%%%%%%%%%%%%%%%%%%%%%%%%%%%%%%%%%%%%%%%
%%%%%%%%%%%%%%%%%%%%%%%%%%%%%%%%%%%%%%%%%%%%%%%%%%%%%%%%%
%\subsection{\large The Factorized Contributions}
\section{\large The Factorized Contributions}
\setcounter{equation}{0} \lbl{fact}

\noi
Let us first discuss the factorized contributions from the $Q_6$
operator. The operators $(\bar{s}_{L}q_{R})$ and
$(\bar{q}_{R}d_{L})$ are Noether densities of the QCD Lagrangian in
Eqs.~\rf{QCDL} and
\rf{dirac}:

{\setl
\bea
(\bar{s}_{L}q_{R}) & = & \bar{q}(x)\lambda_{3j}\gR
q(x)\equiv
D_{\bar{s}q}(x)=-\frac{\delta\cL_{\QCD}(x)}{\delta\cM^{\dagger}(x)_{3j}}
\\ (\bar{q}_{L}d_{R}) & = & \bar{q}(x)\lambda_{j2}\gL
q(x)\equiv
G_{\bar{q}d}(x)=-\frac{\delta\cL_{\QCD}(x)}{\delta\cM(x)_{j2}}\,.
\eea}

\noi
On the other hand, the QCD effective Lagrangian which
describes the strong interactions of Goldstone particles at low
energies is given, in $SU(3)_L\times SU(3)_R$,  by  a string
of terms

{\setl
\bea\lbl{chlag}
\cL_{\chi}& = &\frac{1}{4}F_{0}^2\left\{ \tr
(D_{\mu}UD^{\mu}U^{\dagger})+ 2B\tr (\cM U^{\dagger}+U\cM^{\dagger})
\right\} \nn\\ & & +2B
L_{5}\tr[D_{\mu}U^{\dagger}D^{\mu}U(\cM^{\dagger}U+U^{\dagger}\cM)]
+\cdots
\nn\\ & &  +L_{3}\tr(D_{\mu}U^{\dagger}D^{\mu}U
D_{\nu}U^{\dagger}D^{\nu}U) + iL_{9}\tr
(F_{R}^{\mu\nu}D_{\mu}UD_{\nu}U^{\dagger}+F_{L}^{\mu\nu}
D_{\nu}U^{\dagger}D_{\nu}U)+\cdots\,,
\eea}

\noi
where the first line gives the well known $\cO(p^2)$ terms, and  only
those terms of the
$\cO(p^4)$ Lagrangian~\cite{GL85} which will be needed in this paper
have been explicitly written out. Notice that we
are considering a framework in which the massive $\eta'$ particle has
been integrated out; while, strictly speaking, in a large--$N_c$
formulation one should consider a $U(3)_L\times U(3)_R$ description of
the chiral Lagrangian with an explicit $\eta_{0}$--singlet Goldstone
particle. We have checked that for the observables discussed in this
paper, and at the level of the $\cO(n_f/N_c)$ approximation that we are
retaining, there is no phenomenological difference between the two
formulations.

The bosonization of the  Noether
densities
$D_{\bar{s}q}(x)$ and
$D_{\bar{q}d}(x)$ can then be readily obtained as variations of the
chiral Lagrangian in Eq.~\rf{chlag} with respect to the same external
sources:

{\setl
\bea
D_{\bar{s}q}(x)
& = & -\frac{\delta\cL_{\chi}(x)}{\delta\cM^{\dagger}(x)_{3j}}
\Ra   2B\tr
\lambda_{3j}\left\{\frac{1}{4}F_{0}^{2}U(x)+
L_{5}UD_{\mu}U^{\dagger}D^{\mu}U+\cdots \right\}+\cO(p^4)\,,\lbl{Dsq}
\\
G_{\bar{q}d}(x) & = & -\frac{\delta\cL_{\chi}(x)}{\delta\cM\ (x)_{j2}}
 \Ra  2B\tr
\lambda_{j2}\left\{\frac{1}{4}F_{0}^{2}U^{\dagger}(x)+
L_{5}D_{\mu}U^{\dagger}D^{\mu}U U^{\dagger}+\cdots
\right\}+\cO(p^4)\, ,\lbl{Dqd} \eea}

\noi
where, among the $\cO(p^2)$
contributions, we have only written the term induced by the
$L_{5}$ coupling of the $\cO(p^4)$ chiral Lagrangian because it
is the only term which contributes to the factorized bosonization
of the $Q_6$ operator. Indeed, although the two densities
$D_{\bar{s}q}(x)$ and $D_{\bar{q}d}(x)$ start with terms of
$\cO(p^0)$, their product --because of the fact that the $U$
matrix is unitary-- only starts at $\cO(p^2)$ in the chiral
expansion and, because of the flavour structure,  this product
can only depend on the $L_{5}$ coupling.

On the other hand, the factorized contribution from the $Q_{4}$ operator
is rather straightforward. With

{\setl
\bea
(\bar{s}_{L}\gamma^{\mu}q_{L}) & = &
\bar{q}(x)\lambda_{3j}\gamma^{\mu}\gL q(x)\equiv
L_{\bar{s}q}^{\mu}(x)=\frac{\delta\cL_{\QCD}(x)}{\delta
l_{\mu}(x)_{3j}}\,,
\\ (\bar{q}_{L}\gamma^{\mu}d_{L}) & = &
\bar{q}(x)\lambda_{j2}\gamma^{\mu}\gL q(x)\equiv
L_{\bar{q}d}^{\mu}(x)=\frac{\delta\cL_{\QCD}(x)}{\delta
l_{\mu}(x)_{j2}} \,,
\eea}

\noi
the bosonization of these left--current densities, to lowest order
in the chiral expansion starts at $\cO(p)$ and are given by the
expressions:

{\setl
\bea
L_{\bar{s}q}^{\mu}(x) & \Ra &  \frac{i}{2} F_{0}^2
\tr\left[\lambda_{3j}(D^{\mu}U^{\dagger})U
\right]+\cO(p^3) \lbl{Lsq}
\\ L_{\bar{q}d}^{\mu}(x) & \Ra &  \frac{i}{2} F_{0}^2
\tr\left[\lambda_{j2}(D^{\mu}U^{\dagger})U
\right]+\cO(p^3)\,. \lbl{Lqd}
\eea}

The contribution to the coupling constant $\g_{\underline{8}}$
from the {\it factorized} patterns of the $Q_4$ and $Q_6$ penguin
operators can now be readily obtained from the definitions in
Eqs.~\rf{effH}, \rf{ewcl} and the results in Eqs.~\rf{Dsq},
\rf{Dqd} and \rf{Lsq}, \rf{Lqd}. One thus obtains the well known
result~\footnote{See e.g.~refs.~\cite{ChFG86,BBG,PdeR91} where earlier
references can also be found.}
\be\lbl{g8fact}
g_{\underline{8}}\big\vert^{\mbox{\rm {\scriptsize
factorized}}}_{Q6,Q4}=C_{6}(\mu)\left[-16L_{5}\frac{\stern^2}{F_{0}^2}
\frac{1}{F_{0}^4}\right]+C_{4}(\mu){\bf 1}\,. \ee where we have
used the fact that
\be
B=\frac{1}{F_{0}^2}\vert\stern\vert\,.
\ee
It is well
known that in this approximation,  {\it polychromatic penguins don't
fly,} as emphasized in ref.~\cite{ChFG86},

\vspace*{0.3cm}

%%%%%%%%%%%%%%%%%%%%%%%%%%%%%%%%%%%%%%%%%%%%%%%%%%%%%%%%%
%%%%%%%%%%%%%%%%%%%%%%%%%%%%%%%%%%%%%%%%%%%%%%%%%%%%%%%%%
%\subsection{\large Beyond Factorization}
\section{\large Beyond Factorization}
\setcounter{equation}{0} \lbl{unfact}

\noi Because of the chiral structure of the lowest order electroweak Lagrangian in
Eq.~\rf{ewcl}, we have to expand the Dirac propagators in the {\it unfactorized patterns}
in Eqs.~\rf{Q4} and \rf{Q6} up to two chiral powers in the external sources. There is,
however, a certain arbitrariness in the choice of the external sources to expand. The
simplest (and recommended) choice is the one which makes it explicit that \underline{all}
the perturbative QCD (pQCD) short--distance contributions are factored out in the Wilson
coefficients $C_{i}(\mu)$. Since the operators $Q_{4}$ and $Q_{6}$ transform like
$\underline{8}_{L}\times \underline{1}_{R}$ operators under chiral rotations, expanding
the propagators in right--handed $r_{\mu}$ sources will necessarily bring in Green's
functions with extra $R_{\mu}\equiv \bar{q}\gamma_{\mu}\gR q$ currents only, and there is
no way then that the product of these $(\underline{1}_{L}\times\underline{8}_{R})$
$R_{\mu}$  operators with the initial $(\underline{8}_{L}\times \underline{1}_{R})$
operator can produce a $\underline{1}_{L}\times \underline{1}_{R}$ term, which ensures no
mixing with possible pQCD contributions in the long--distance evaluation. Technically,
this can be accomplished by an appropriate {\it``tout \`a droite''} rotation~\footnote{In
the french cycling jargon, ``tout \`a droite''  means to put the front and rear gears so
as to reach maximum speed (the biggest front plateau with the smallest rear socket). This
is what we recommend to do here with the Dirac operator! } of the quarks fields so that
in the rotated basis the right-handed quark field has absorbed all the Goldstone degrees
of freedom and the Dirac operator has as a chiral vector connection the external
$l_{\mu}$ field \underline{only}. One way to implement this is as follows: define $Q_{L}$
and $Q_{R}$ quark fields, such that \be Q_{L}(x)=\xi_{L}(x)q_{L}(x)\quad\annd\quad
Q_{R}(x)=\xi_{R}(x)q_{R}(x)\,, \ee where $\xi_{L}(x)$ and $\xi_{R}(x)$ are left and right
$SU(3)_{L}\times SU(3)_{R}/SU(3)_V$ coset representatives , with
$\xi_{R}(x)\xi_{L}^{\dagger}(x)=U(x)$. Then, choose a gauge where \be
\xi_{L}(x)=1\,,\quad\mbox{\rm while}\quad \xi_{R}(x)=U(x)\,. \ee The Dirac operator in
this {\it ``tout \`a droite''} rotated basis is rather simple \be\lbl{diractag} D_{\chi}
= \gamma^{\mu}\left(\partial_{\mu}+iG_{\mu}-il_{\mu}\frac{1-\gamma_5}{2}\right)+
U^{\dagger}D_{\mu}U \gamma^{\mu}\gR +iU^{\dagger}\cM\ \gL+i\cM^{\dagger}U\ \gR\,. \ee The
chiral expansion, in the chiral limit where $\cM=\cM^{\dagger}=0$, can then be made in
powers of the external source which collects the full Goldstone structure; i.e., the term
$U^{\dagger}D_{\mu}U \gamma^{\mu}\gR $ in Eq.~\rf{diractag}.

%%%%%%%%%%%%%%%
%%%%%%%%%%%%%%%
\subsection{\sc\large Determination of the unfactorized Green functions}
%\subsection{\sc\large The Unfactorized Contributions in the
%$1/N_c$ Expansion}
%\setcounter{equation}{0}
\lbl{unfactW}

\noi
As already mentioned, we are only considering in this work the
bosonization of the {\it unfactorized} terms in Eqs.~\rf{Q4} and \rf{Q6}
of
$\cO(N_c^2\frac{n_f}{N_c})$ in the $1/N_c$ expansion, and in the
chiral limit. As we shall see, the Physics features which
already emerge at that level deserve attention. It is possible,
however, to extend the calculation in the chiral limit to
the rest of the $\cO(N_c^2\frac{1}{N_c})$ contributions, which are not
enhanced by a
$n_f$ factor, something which we plan to do in the near future.

%%%%%%%%%%%% Figure 2 %%%%%%%%%%
\vskip 2pc
\centerline{\epsfbox{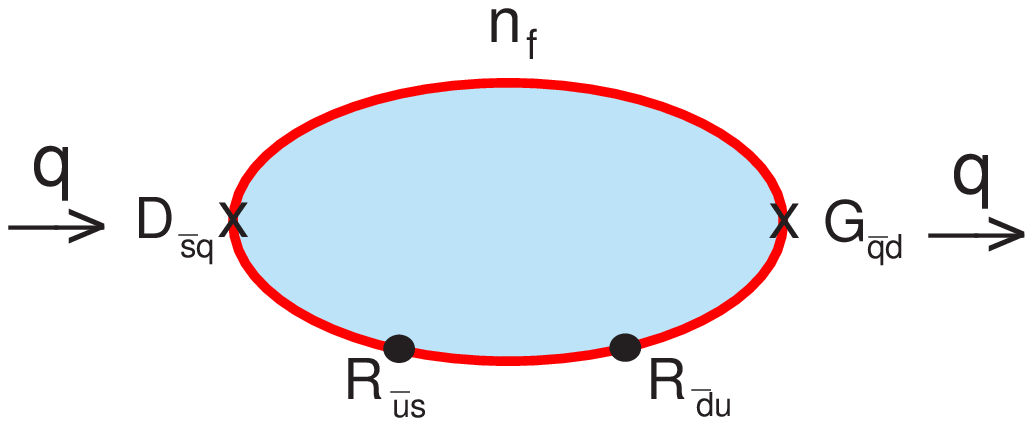}}\vspace*{0.5cm}
%\centerline{
{\bf Fig.~2} {\it ``Eye''--like configurations contributing to
the unfactorized bosonization of the $Q_6$ operator. The solid
dots in the quark propagator correspond to external right--handed
sources of {\it soft} quark currents.}
%}
\vskip 2pc
%%%%%%%%%%%%%%%%%%%%%%%%%%%%%%

Let us discuss the
case of the $Q_6$ operator in some detail. The only configurations which
can contribute in the approximation that we are considering, are the so
called ``eye''--like configurations~\cite{BDSPW}, where the two right--handed
sources come out from the same propagator in Eq.~\rf{Q6}, as illustrated in
Fig.~2. They produce the
functional integral~\footnote{The contribution from the term where the
first propagator in Eq.~\rf{Q6} is expanded instead of the second one, is
modulated by the flavour factor
$\sum_{j}\lambda_{j2}\lambda_{3j}=0$, and therefore it vanishes.}

{\setl
\bea
\lefteqn{\langle Q_6 \rangle\vert_{\mbox{\rm\tiny
eye.}}=
8\ n_{f}\int\frac{d^4 q}{(2\pi)^4}\int d^4y\
\Tr(x\vert\gR\frac{i}{\psls+\frac{\qsls}{2}}\vert
y)\times} \nn \\ & & (y\vert\lambda_{32}\gL
\frac{i}{\psls-\frac{\qsls}{2}}\left(U^{\dagger}D_{\alpha}U
\gamma^{\alpha}\gR\right)\frac{i}{\psls-\frac{\qsls}{2}}
\left(U^{\dagger}D_{\beta}U
\gamma^{\beta}\gR\right)\frac{i}{\psls-\frac{\qsls}{2}}\vert
x)\,,
\eea}

\noi which gives a term with the required chiral structure: $ 8
n_f\times\tr\lambda_{32}U^{\dagger}D_{\alpha}U U^{\dagger}D_{\beta}U =-8
n_f\times\tr\lambda_{32}D_{\alpha}U^{\dagger}D_{\beta}U\,, $ modulated by the integral
(recall Eq.~\rf{Gprop}) \be\lbl{qcdint6} \int\ \frac{d^4q}{(2\pi)^4}\int d^4y
\Tr(x\vert\gR\frac{i}{\psls+\frac{\qsls}{2}}\vert y)(y\vert\gL\frac{i}
{\psls-\frac{\qsls}{2}}\gamma^{\alpha}\gR \frac{i}
{\psls-\frac{\qsls}{2}}\gamma^{\beta}\gR \frac{i} {\psls-\frac{\qsls}{2}}\vert x)\,. \ee
The integral in question can be related to a well defined QCD Green's function, which is
the connected Green's function \be\lbl{DGRR} \cW_{DGRR}^{~~~~\alpha\beta}(q,k)\!=\!
\lim_{k\ra 0}\ i^3 \int d^4x d^4y d^4z e^{iq\cdot x} e^{ik\cdot y} e^{-ik\cdot z}\langle
0\vert T\{ D_{\bar{s}q}(x)G_{\bar{q}d}(0)
R_{\bar{d}u}^{\alpha}(y)R_{\bar{u}s}^{\beta}(z)\}\vert 0\rangle\vert_{\mbox{\rm\tiny
conn.}}\,, \ee where (no summation over the flavour index $q$ in Eq.~\rf{DGRR}) \be
D_{\bar{s}q}(x)=\bar{q}(x)\lambda_{3j}\gR q(x)\,,\qquad
G_{\bar{q}d}(0)=\bar{q}(0)\lambda_{j2}\gL {q}(0)\,, \ee and \be
R_{\bar{d}u}^{\alpha}(y)=\bar{q}(y)\lambda_{21}\gamma^{\alpha}\gR q(y)\,,\qquad
R_{\bar{u}s}^{\beta}(z)=\bar{q}(z)\lambda_{13}\gamma^{\beta}\gR q(z)\,. \ee This Green's
function can be viewed as a two--point function of the two $D_{\bar{s}q}(x)$ and
$G_{\bar{q}d}(0)$ operators which define $Q_{6}$, with two soft insertions of the
composite operators $R_{\bar{d}u}^{\alpha}(y)$ and $R_{\bar{u}s}^{\beta}(z)$. The
relation to the integral in Eq.~\rf{qcdint6}, and hence to $\langle Q_6
\rangle\vert_{\mbox{\rm\tiny eye.}}$, is the following \be \langle Q_6
\rangle\vert_{\mbox{\rm\tiny eye.}}=-\tr\lambda_{32}D_{\alpha}U^{\dagger}D_{\beta}U\times
8n_f\times\ \left(\int\frac{d^4q}{(2\pi)^4}\cW_{DGRR}^{~~~~\alpha\beta}(q,k)
\right)\Bigg\vert_{g^{\alpha\beta}}\,. \ee with the integral restricted to the term
proportional to $g^{\alpha\beta}$. In fact, this term appears naturally, when performing
the integral over the solid angle $d\Omega_{q}$, which has the form ($Q^2\equiv -q^2$):
\be\lbl{intDGRR} \int d\Omega_{q}\ \cW_{DGRR}^{~~~~\alpha\beta}(q,k)=
\left(\frac{k^{\alpha}k^{\beta}}{ k^2}-g^{\alpha\beta}\right)\ \cW_{DGRR}(Q^2)\,, \ee
where the transversality in the four--vector $k$ follows from Current Algebra Ward
identities. We are still left with  an integral of the invariant function
$\cW_{DGRR}(Q^2)$  over the full euclidean range ($0\le Q^2\le\infty$) which has to be
done in the same {\it renormalization scheme} which has been adopted when doing the
calculation of the short--distance Wilson coefficient $C_{6}(\mu)$ in Eq.~\rf{effH}, i.e.
in the $\msb$ scheme.

The case of the unfactorized bosonization of the $Q_{4}$ operator
is rather similar. To $\cO(p^2)$ it is governed by a connected
Green's function  which can be viewed as a two--point function of
the two $L_{\bar{s}q}^{\mu}(x)$ and $L_{\bar{q}d}^{\nu}(0)$
operators which define $Q_{4}$, with two soft insertions of the
composite operators $R_{\bar{d}u}^{\alpha}(y)$ and
$R_{\bar{u}s}^{\beta}(z)$; i.e.,
\be\lbl{Q4UB} \langle Q_4
\rangle\vert_{\mbox{\rm\tiny
eye.}}=\tr\lambda_{32}D_{\alpha}U^{\dagger}D_{\beta}U\times 4n_f\times
\left(g_{\mu\nu}
\int\frac{d^4q}{(2\pi)^4}\cW_{LLRR}^{\,\mu\,\nu\,\alpha\,\beta}(q,k)
\right)\Bigg\vert_{g^{\alpha\beta}}\,.
\ee
where here
\be\lbl{LLRR}
\cW_{LLRR}^{\,\mu\,\nu\,\alpha\,\beta}(q,k)\!=\!\lim_{k\ra 0}\ i^3
\int d^4x d^4y d^4z e^{iq\cdot x} e^{ik\cdot y}
e^{-ik\cdot z}\langle 0\vert T\{
L_{\bar{s}q}^{\mu}(x)L_{\bar{q}d}^{\nu}(0)
R_{\bar{d}u}^{\alpha}(y)R_{\bar{u}s}^{\beta}(z)\}\vert
0\rangle\vert_{\mbox{\rm\tiny conn.}}\,,
\ee
with no summation over the flavour $q$ index in the currents.
The Ward identities again ensure that the integral over
the solid angle in Eq.~\rf{Q4UB} also depends on one invariant
function only,
\be\lbl{intLLRR} g_{\mu\nu}\int d\Omega_{q}\
\cW_{LLRR}^{\,\mu\,\nu\,\alpha\,\beta}(q,k)=
\left(\frac{k^{\alpha}k^{\beta}}{ k^2}-g^{\alpha\beta}\right)\
\cW_{LLRR}(Q^2)\,,
\ee
and the term proportional to
$g^{\alpha\beta}$ appears then naturally, as well.

It is now possible to express the contribution to the coupling constant
$g_{\underline{8}}$ of the electroweak chiral Lagrangian in
Eq.~\rf{ewcl} from the {\it factorized patterns} of the $Q_4$ and $Q_6$
penguin operators to $\cO(N_c^2)$, as well as from their {\it unfactorized
patterns} of $\cO(N_c^2\frac{n_f}{N_c})$, which appear as integrals  of
the invariant functions $\cW_{DGRR}(Q^2)$ and $\cW_{LLRR}(Q^2)$, in the
following way:

{\setl
\bea\lbl{g8int}
\g_{\underline{8}}\vert_{Q_4,Q_6} & = &
C_{6}(\mu)\left\{\!-16L_{5}\frac{\stern^2}{F_{0}^6}\!+\!
\frac{8n_f}{16\pi^2
F_{0}^4}\left[\frac{(4\pi\mu^2)^{\epsilon/2}}{\Gamma(2-\epsilon/2)}
\!\int_{0}^{\infty}\!\!dQ^2
(Q^{2})^{1-\epsilon/2}\cW_{DGRR}(Q^2)\!\right]_{\msb}\right\} \nn
\\
 & + & C_{4}(\mu)\left\{1-\frac{4n_f}{16\pi^2
F_{0}^4}\left[\frac{(4\pi\mu^2)^{\epsilon/2}}{\Gamma(2-\epsilon/2)}
\int_{0}^{\infty} dQ^2
(Q^{2})^{1-\epsilon/2}\cW_{LLRR}(Q^2)\right]_{\msb}
\right\}\,.
\eea}

\noi
Notice that the integrals over $Q^2$ are
$\msb$ renormalized; i.e., poles in $1/\epsilon$ as well as the
constant $\log 4\pi-\gamma_{E}$ are removed. Let us remember,
however, that the scale dependence of the renormalized constant
$L_{5}(\mu)$ in $\chi$PT is \be L_{5}(\mu)=L_{5}(M_{\rho})-
\frac{\Gamma_{5}}{16\pi^2}
\log\frac{\mu}{M_\rho}\,,\quad\with\quad \Gamma_{5}=\frac{n_f}{8}\,,
\ee
and coupling constants in $\chi$PT have been renormalized in
the Bern $\msb$ scheme where  poles in $1/\epsilon$, as well as
the constant $\log{4\pi}-\gamma_{E} +1$, are removed. This
difference of $\msb$ renormalization schemes has to be taken into
account consistently in order to exhibit the overall $\mu$ scale
cancellation between the short-- and long--distance contributions, at the
order of approximation that we are working in the $1/N_c$ expansion.
Here we have chosen to renormalize all the UV divergences with a single scale
$\mu^\epsilon$. Therefore, in Eq.~\rf{g8int} there is a
$\mu$--scale dependence both in parameters such as
$L_5(\mu)$ and $\stern(\mu)$, as well as in the short--distance Wilson
coefficients, such as $C_{6,4}(\mu)$. There is,
however, freedom to renormalize $L_5$  with a different scale, which could
be called $\nu_{\mbox{\rm\tiny chiral}}$ as conventionally done in $\chi$PT.
That this is possible is due to the fact that the
renormalization of $L_5$ does not mix with that of the quark--condensate,
or the Wilson coefficient.  Notice also that, in the $\msb$
renormalization scheme which we are using, where quark masses (and hence the
pseudoscalar masses) are set to zero, there are no chiral loop
contributions to the factorized term in Eq.~\rf{g8int}. The fact that all the
UV scale $\mu$ dependence cancels out in the end in the coupling constant
$g_{\underline{8}}$ is a nontrivial check that the interplay between
short-- and long--distances has been brought under control.

%%%%%%%%%%%%%%%
%%%%%%%%%%%%%%%
\subsection{\sc\large Why one could expect large unfactorized
contributions}
\lbl{argum}

At this level, we think it necessary to have a short discussion
explaining the relative importance of the {\it factorized} versus {\it
unfactorized} contributions in Eq.~(\ref{eq:g8int}). It has long been
recognized that the calculation of the coupling constant
$\g_{\underline{8}}$ is a multi--scale problem. Already at
short distances, the perturbative running of the Wilson coefficients has
to deal with several different scales from $M_W$ down to the charm mass.
However, it is perhaps not so much emphasized that, even in the
long--distance regime, there are also two distinctive scales playing a
role in Eq. (\ref{eq:g8int}). Although in QCD all scales are ultimately
related to $\Lambda_{\overline{\mathrm{MS}}}$, it is a fact of
life that $F_0$ is much smaller than a typical hadronic scale, of the size
of an average resonance mass, $M_{R}\sim 1$ GeV. Also the typical scale
defined by a QCD vacuum condensate\footnote{We are thinking here of
something like
$\stern^{1/3}$ at $\mu=1$ GeV.}, $\Lambda_c$, is known to be smaller
than $M_R$. One then has that, roughly, $F_0\ll \Lambda_c\ll
M_R$.
Because
$F_0^2$ is of $\cO(N_c)$ while $M_R^2$ is of $\cO(1)$ a naive
interpretation of the large--$N_c$ counting could lead one to conclude
that
$F_0^2\gg M_R^2$, which is obviously wrong. It is this hierarchy of
scales which is at the origin of why the {\it unfactorized} contribution,
although of
$\mathcal{O}(N_c)$, may be as large as --or sometimes larger
than-- the {\it factorized} one, even though the latter is formally
of  $\mathcal{O}(N_c^2)$.  For instance, we already found that in the case
of the calculation of
$B_K$ in the chiral limit~\cite{PdeR00,CP03}, the unfactorized
contributions are of the order of 50\%. In the present case of the
$Q_6\,, Q_4$ penguin operators,  the unfactorized contributions actually
turn out to be much larger than the factorized ones, as we shall soon
see. We conclude that in the case of four--quark operators,  a naive
interpretation of the
$1/N_c$ expansion may be misleading and has to be done
{\it after} the contributions from the different physical
scales have been separated.

In the case we are considering, knowing that (as will be confirmed in the
next section by explicit calculation):
\be\lbl{integrals}
\int_{0}^{\infty} dQ^2 Q^2 \ \mathcal{W}_{DGRR}(Q^2)\sim B^2F_0^2\,,\quad
\mathrm{and}\quad \int_{0}^{\infty} dQ^2 Q^2\  \mathcal{W}_{LLRR}(Q^2)\sim
F_0^2 M_R^2\,,
\ee
up to logarithmic factors, and that  $L_5\sim F_{0}^2/M_R^2$,
the ratio of the {\it unfactorized} to the {\it factorized}
contributions in Eq. (\ref{eq:g8int}) is given by precisely the ratio
of scales
\be\lbl{message}
\frac{{unfactorized}}{{factorized}}\sim
\frac{M_R^2}{16 \pi^2 F_0^2}\ ,
\ee
which is a number of $\mathcal{O}(1)$, although large--$N_c$
suppressed. This is true for \emph{both} the contributions from
$Q_6$ and $Q_4$ to $g_{\underline{8}}$ in Eq. (\ref{eq:g8int}).

Is this a shortcoming of the large$-N_c$ expansion? Not necessarily.
It is rather that the standard analysis, as  used, e.g., in
ref.~\cite{W79} was only applied to situations in which the two scales
$F_0$ and $M_R$ do not compete for the control of the large--$N_c$
expansion, as it is the case in Eq.~(\ref{eq:g8int}). In fact, we
think that there is nothing specially new about this dynamical effect of
scales. Even the perturbative expansion in QED can be affected by the
presence of two widely separated masses, because they give rise to large
logarithms of their ratio. In the case of the electroweak theory,
Bardeen {\it et al.} in ref.~\cite{BBG} were the first to point out that
the large--$N_c$ expansion is also affected by the presence of the large
value of
$\log M_W/m_{\mathrm{charm}}$. The point that we would like to
emphasize here is that, also in the low--energy region, ratios of
scales such as $M_R^2/(4\pi F_0)^2$ may upset a large--$N_c$
counting done in a naive way. On the contrary, once the
contributions from these two scales have been clearly separated,
it should be safer applying a large--$N_c$ argument. For instance,
one might fear that since the unfactorized term is larger than the
factorized one, the whole expansion breaks down and the next
subleading terms will be even bigger. Although a definitive
answer to this point can only come from the solution to QCD in
the large--$N_c$ expansion, which is not known, we would like to
argue that there is no reason to expect this breakdown.
Subleading terms to those of Eq. (\ref{eq:g8int}) are expected to give
contributions of the order of the width of the resonances involved, or of
the violation of the Zweig rule. There is presently no evidence against
these effects being reasonably small except, perhaps, in some very
specific cases  where the effect of the  strong
Goldstone--Goldstone interaction in the
$J=I=0$ channel appears, which may deserve special attention and which
we plan to study elsewhere.

%%%%%%%%%%%%%%
%%%%%%%%%%%%%%
\subsection{\sc\large Long and short distance constraints on the
functions $\cW_{DGRR}(Q^2)$ and
$\cW_{LLRR}(Q^2)$ }
\lbl{unfactConstr}

\noi The low--$Q^2$ behaviour of these functions is governed by
$\chi$PT. An explicit calculation gives the results:

{\setl
\bea\lbl{chDG}
\lim_{Q^2\ra 0}\cW_{DGRR}(Q^2) &= & \frac{1}{8}\times
\frac{BF_{0}}{Q^2}\times
\frac{BF_{0}}{Q^2}+
 \left(
-L_{5}+\frac{5}{2}L_{3}\right)\frac{B^2}{Q^2}+\cO(\mbox{\rm\small
Cte.})\,, \\
\lim_{Q^2\ra 0}\cW_{LLRR}(Q^2) &=&
-\frac{3}{8}\frac{F_{0}^2}{Q^2}
+(-\frac{15}{2}L_{3}+\frac{3}{2}L_9)+\cO(Q^2)\,. \lbl{chLL}
\eea}

\noi
The double pole contribution to $\cW_{DGRR}(Q^2)$, the $L_5$
contribution to $\cW_{DGRR}(Q^2)$ and the simple pole contribution to
$\cW_{LLRR}(Q^2)$
agree
with the result first obtained in refs.~\cite{HKPSB98,HKPS99}
and~\cite{HKS99} respectively~\footnote{The $L_3$ contribution to
Eq.~\rf{chDG} was omitted in ref.~\cite{HKPS99} for reasons  which turn
out to be incorrect.}. To our knowledge, the other terms have not been
calculated before.

On the other hand, the high--$Q^2$ behaviour
of the functions $\cW_{DGRR}(Q^2)$ and
$\cW_{LLRR}(Q^2)$ is governed by the operator product expansion of the
$D\bigotimes G$ density currents in Eq.~\rf{DGRR} and the $L\bigotimes L$
currents in Eq.~\rf{LLRR}, with the results:

{\setl \bea\lbl{opecoeff1} \lim_{Q^2\ra\infty}\cW_{DGRR}(Q^2) &
= & -\frac{1}{6}\
\pi^2\frac{\als}{\pi}\frac{F_{0}^4}{Q^4}+\frac{8}{3}\ \pi^2
\frac{\als}{\pi}
\frac{\stern^2}{F_{0}^2}\frac{L_5}{Q^4}+\cO[\epsilon\frac{1}{Q^4}]\,, \\
\lim_{Q^2\ra\infty} \cW_{LLRR}(Q^2) & = & +\frac{1}{3}\
\pi^2\frac{\als}{\pi}\frac{F_{0}^4}{Q^4}-\frac{16}{3}\pi^2
\frac{\als}{\pi}
\frac{\stern^2}{F_{0}^2}\frac{L_5}{Q^4}+\cO[\epsilon\frac{1}{Q^4}]
\lbl{opecoeff2}\,.
\eea}

\noi We remark that in Eqs. \rf{chDG}-\rf{opecoeff2} the chiral couplings $L_{5,9}$ are
the ones corresponding to leading order at large-$N_c$ and, therefore, do not run with
the scale.

In this work we shall content ourselves with only the leading--log approximation. This
allows us to do away with the evaluation of  the terms of $\cO[\epsilon\frac{1}{Q^4}]$ in
the OPE, which means that we shall not be able to specify {\it scheme} dependences in our
$\msb$--renormalization calculation of the long--distance effects.

In large--$N_c$ QCD, the Green's functions $\cW_{DGRR}(Q^2)$ and
$\cW_{LLRR}(Q^2)$ are meromorphic functions in the $Q^2$ variable.
The most general structure they can have is:

{\setl
\bea\lbl{largeNDGRR}
Q^2\cW_{DGRR}(Q^2)
& = & \frac{1}{8}\frac{(BF_{0})^2}{Q^2}+ \nn\\
&  & \muhad^4\sum_{i}
\frac{\alpha_{i}}{Q^2+M_{i}^2}+
\muhad^6\sum_{i}
\frac{\beta_{i}}{(Q^2+M_{i}^2)^2}+
\muhad^8\sum_{i}
\frac{\gamma_{i}}{(Q^2+M_{i}^2)^3}\,,  \lbl{hadfase}\\\lbl{largeNLLRR}
Q^2\cW_{LLRR}(Q^2)
& = &
\muhad^4\sum_{j}
\frac{\alpha'_{j}}{Q^2+M_{j}^2}
+\muhad^6\sum_{i}
\frac{\beta'_{j}}{(Q^2+M_{j}^2)^2}+
\muhad^8\sum_{i}
\frac{\gamma'_{j}}{(Q^2+M_{j}^2)^3}\,, \lbl{hadfaseLL}
\eea}

\noi where $\muhad$ is an arbitrary hadronic mass scale,  so as to make the residues
$\alpha_{i}$, $\beta_{i}$, $\gamma_{i}$ and $\alpha'_{j}$, $\beta'_{j}$, $\gamma'_{j}$
dimensionless. This scale can be interpreted as the scale at which the four--quark
Lagrangian in Eq.~\rf{effH} is matched onto the chiral Lagrangian of  Eq.~\rf{ewcl}, i.e.
the scale at which the meson resonances are integrated out. Physical results, however, do
not depend on the particular choice of the scale $\muhad$, except for higher order terms.
The sums are, in principle, extended to an infinite number of states, as illustrated in
Fig.~3. In practice, it is useful to start with the {\it minimal hadronic
approximation}~\footnote{See e.g. ref.~\cite{deR02} and references therein for details}
(MHA) where the number of non--Goldstone states is limited to a minimum number of lowest
mass states, with appropriate quantum numbers; the minimum number which is necessary to
satisfy the leading OPE constraints. In our case, the MHA we shall adopt is the one with
a $1^{-}$ vector pole of mass $M_V$ and a $0^{+}$ scalar pole of mass $M_S$.

%%%%%%%%%%%% Figure 3 %%%%%%%%%%
\vskip 2pc \centerline{\epsfbox{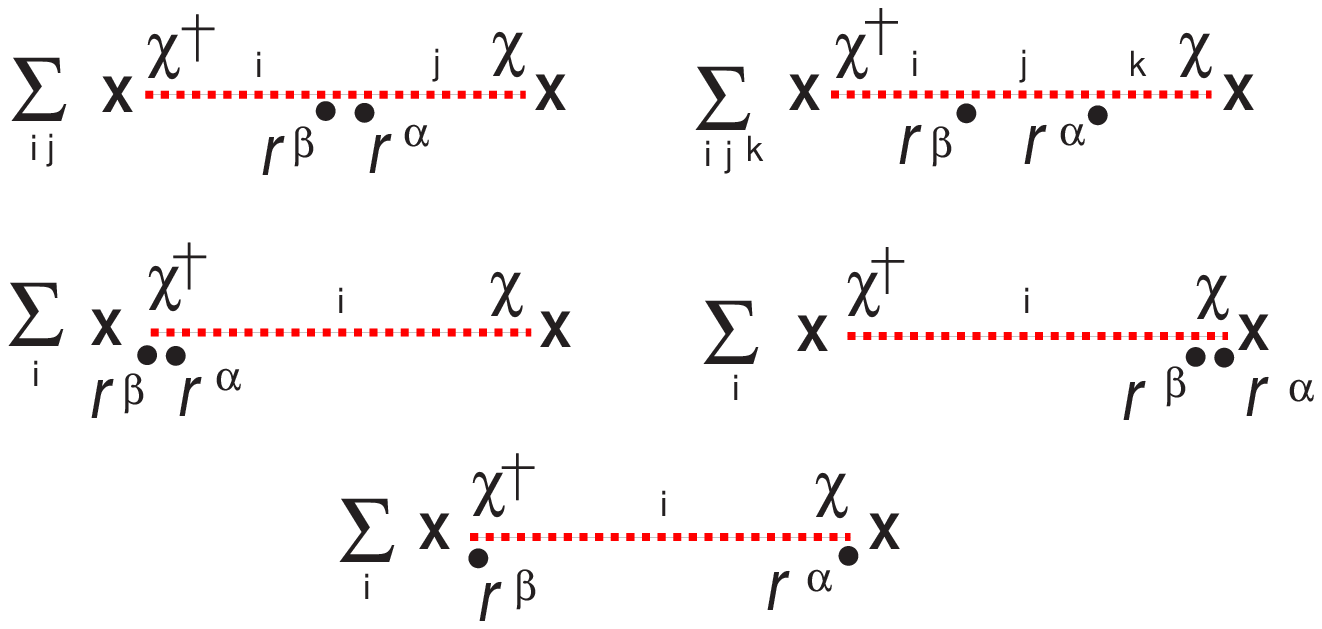}}\vspace*{0.5cm}
%\centerline{
{\bf Fig.~3} {\it Different types of hadronic tree diagrams which can
contribute to the Green's functions $Q^2\cW_{DGRR}(Q^2)$ and
$Q^2\cW_{LLRR}(Q^2)$ (replacing $\chi\ra l$) in the large--$N_c$ limit. The
indices $i$, $j$, $k$ label the possible poles.}
%}
\vskip 2pc
%%%%%%%%%%%%%%%%%%%%%%%%%%%%%%

\noi
With
the notation ($M_V=770~\MeV$ and $M_S\simeq 1~\GeV$)
\be z=\frac{Q^2}{\muhad^2}\,,\quad
\rho_V=\frac{M_V^2}{\muhad^2}\quad\annd\quad
\rho_S=\frac{M_S^2}{\muhad^2}\,,
\ee
\noi
the MHA parameterizations are,
\be\lbl{mha6} z\cW_{DGRR}^{(\MHA)}(z
\muhad^2)=\frac{1}{8}\left(\frac{BF_0}{\muhad^2}\right)^2\frac{1}{z}
+\frac{\alpha_V}{z+\rho_V}+\frac{\alpha_S}{z+\rho_S}\,,
\ee
and
\be\lbl{mha4} z\cW_{LLRR}^{(\MHA)}(z\muhad^2)=
\frac{\alpha'_V}{z+\rho_V}+\frac{\alpha'_S}{z+\rho_S}\,,
\ee
with
the residues $\alpha_V$, $\alpha_S$, $\alpha'_V$ and $\alpha'_S$
solutions of the system of equations defined by the
short--distance constraints

{\setl
\bea\lbl{sdDG}
\frac{1}{8}(BF_{0})^2+\muhad^4\sum_{i=V,S}\alpha_{i} & = &
-\frac{1}{6}\
\pi^2\frac{\als}{\pi}F_{0}^4+\frac{8}{3}\ \pi^2 \frac{\als}{\pi}
\frac{\stern^2}{F_{0}^2}L_5\,, \lbl{OPEDG1}\\
\muhad^4\sum_{j=V,S}\alpha'_{j} & = & +\frac{1}{3}\
\pi^2\frac{\als}{\pi}F_{0}^4-\frac{16}{3}\pi^2 \frac{\als}{\pi}
\frac{\stern^2}{F_{0}^2}L_5\,, \lbl{OPELL1}
\eea}

\noi
and the long--distance constraints

{\setl
\bea
\muhad^2 \sum_{i=V,S} \frac{1}{\rho_i}\ \alpha_i &=& -\left(
L_5 -
\frac{5}{2} L_3 \right) B^2\,, \lbl{p4DG}\\
\muhad^2 \sum_{j=V,S} \frac{1}{\rho_j}\alpha'_j &=& -\frac{3}{8}
F^2_0\,.
\lbl{p2LL}
\eea}

%%%%%%%%%%%% Figure 4 %%%%%%%%%%
\vskip 2pc
\centerline{\epsfbox{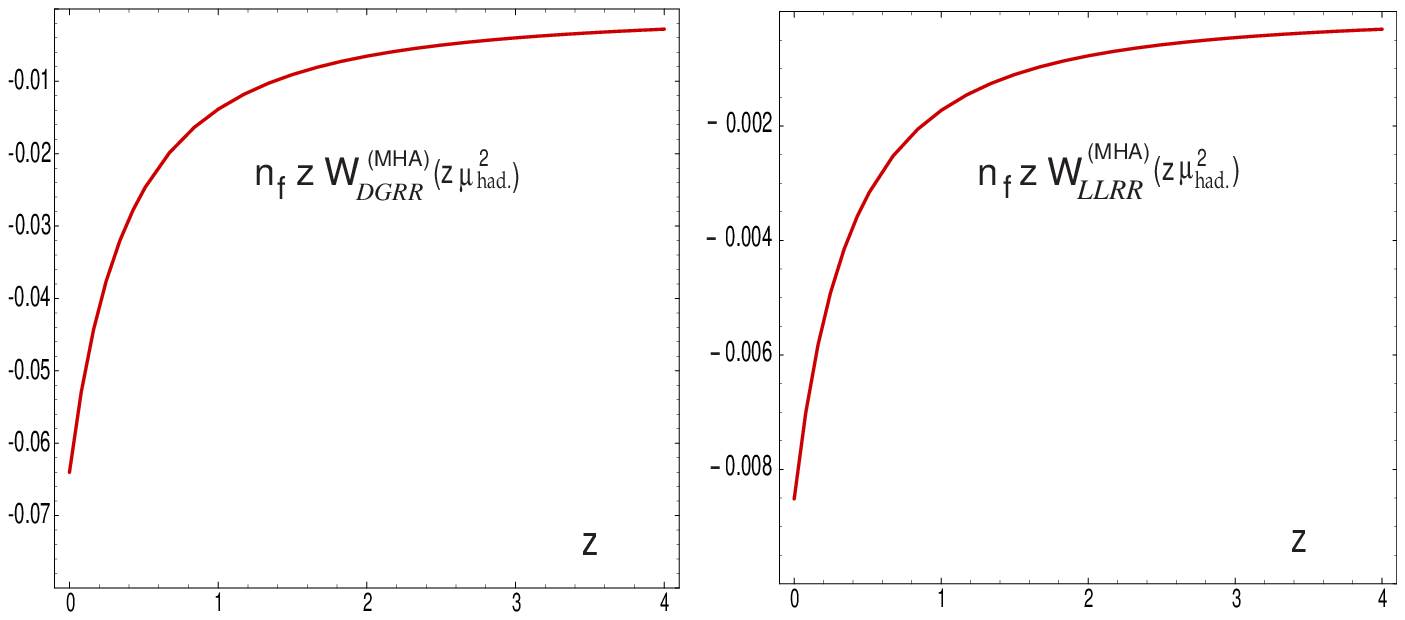}}\vspace*{0.5cm}
{\bf Fig.~4} {\it The functions $z\cW_{DGRR}^{(\MHA)}(z\muhad^2)$ (with
the Goldstone pole removed) and  $z\cW_{LLRR}^{(\MHA)}(z\muhad^2)$ in
Eqs.~\rf{mha6} and \rf{mha4}, versus $z=Q^2/\muhad^2$ for the particular
choice
$\muhad=1~\GeV$ and $\vert\stern\vert^{1/3}=250~\MeV$. The values of the other
input parameters are specified in the Appendix. Notice also the different
vertical scales in the two plots.}

\vskip 2pc
%%%%%%%%%%%%%%%%%%%%%%%%%%%%%%
\vspace{0.7cm}

\noi The shapes of the resulting functions in
Eqs.~(\ref{eq:mha6}) and (\ref{eq:mha4}) are shown in Fig.~4.
A few comments are now in order:

\begin{itemize}

\item
The reason why we are showing the shape of the functions
$zW_{DGRR}^{(\MHA)}(z)$ and $zW_{LLRR}^{(\MHA)}(z)$, instead of
$W_{DGRR}^{(\MHA)}(z)$ and $ W_{LLRR}^{(\MHA)}(z)$ themselves, is that
these are precisely the relevant integrands which appear in
the integrals in Eq.~\rf{g8int}. The pion pole in $z\times
W_{DGRR}^{(\MHA)}(z)$ does not contribute to the ``area'' in dimensional
regularization, and this is why we can remove it from the integrand
with impunity.

\item
In spite of the fact that the Goldstone pole (i.e. the term proportional to
$(BF_0)^2$in Eq.~\rf{mha6}) does not contribute to the $Q^2$
integral in dimensional regularization, the sum of the residues of the
hadronic simple poles
$\sum_{i}\alpha_{i}$ is constrained by the {\it
residue} of this Goldstone pole through the OPE, as shown in
Eq.~\rf{sdDG}; and in fact, it is this term which largely dominates the
short--distance constraint. This is the subtle way in which the lowest order
chiral Lagrangian, which does not contribute at the factorized level,  shows
its presence at the unfactorized level. It is precisely this term which
is responsible for the renormalization of $L_5$.

\item
The residue of the single Goldstone pole contribution to  $\cW_{DGRR}$ is
fixed by a combination of couplings of the $\cO (p^4)$ chiral Lagrangian,
as seen in Eq.~\rf{chDG}. The same residue provides a constraint on the
sum $\sum_{i}\frac{\alpha_{i}}{\rho_i}$ of the hadronic parameters, as
seen in Eq.~\rf{p4DG}.

\item
Numerically, the chiral factor $L_5-\frac{5}{2}L_3\simeq
10^{-2}$ which appears in Eqs.~\rf{chDG} and \rf{p4DG} is rather large, and it
is at the origin of the fact that the {\it unfactorized} contributions will
turn out to be so important, since it is this quantity (times $\stern^2$)
which fixes the value at the origin of the integrand $Q^2\times
W_{DGRR}(Q^2)$ in Eq.~\rf{g8int}.

\item
The corresponding chiral slope factor $-\frac{15}{2}L_3+\frac{3}{2}L_9\simeq
30\times 10^{-3}
$ for
$\cW_{LLRR}(Q^2)$ in Eq.~\rf{chLL} is even larger. However, in this case,
the intercept at the origin is fixed by the $\cO(p^2)$ term
$-\frac{3}{8}F_0^2$ (see Eq.~\rf{p2LL}), which is of a reasonable size.

\end{itemize}

The MHA parameterizations of the functions $\cW_{DGRR}^{(\MHA)}(Q^2)$ and
$\cW_{LLRR}^{(\MHA)}(Q^2)$ may, however, be improved. On the short--distance side, they
only approach their OPE behaviour at rather large $Q^2$ values. On the long--distance
side  the behaviour of $\cW_{LLRR}^{(\MHA)}(Q^2)$ at small $Q^2$ fails to reproduce the
known slope in Eq.~\rf{chLL}. This is why we also went beyond the MHA and considered more
sophisticated parameterizations of the general structure given in Eqs.~\rf{largeNDGRR}
and \rf{largeNLLRR}, demanding that they interpolate smoothly between the known chiral
behaviour and the OPE behaviour. In the case of $\cW_{DGRR}(Q^2)$ we included a vector
simple pole, a scalar simple and double poles, and an excited pseudoscalar pole. For
$\cW_{LLRR}(Q^2)$ we included a scalar simple pole together with simple, double and
triple pole for a vector, an excited vector, and an axial--vector states.  The extra
information to fix the residues of the poles was obtained by demanding that
$\cW_{DGRR}(Q^2)$ and $\cW_{LLRR}(Q^2)$ reproduce the OPE at various points, $Q^2\gtrsim
9~\GeV^2$. This scale was arbitrarily chosen as being low enough to help producing a
smooth interpolation but high enough for the OPE behaviour to be trusted. Typical
examples of the shapes we get for $n_f z\cW_{DGRR}(Q^2)$ and $n_f z\cW_{LLRR}(Q^2)$ are
shown, respectively, in Figs.~5 and 6 below (the thick solid curves). For the purpose of
comparison, we also show in the same plots  the OPE behaviour (the thick dashed curves),
as well as the $\chi$PT slope in the case of  $z\cW_{LLRR}(Q^2)$ (the thin dashed line in
Fig.~6). The slope of $z\cW_{DGRR}(Q^2)$ (with the pion pole removed) could be determined
from the couplings of the $\cO(p^6)$ chiral Lagrangian. The calculation, which is rather
involved, is beyond the scope of this paper.

%%%%%%%%%%%% Figure 5 %%%%%%%%%%
\vskip 2pc
\centerline{\epsfbox{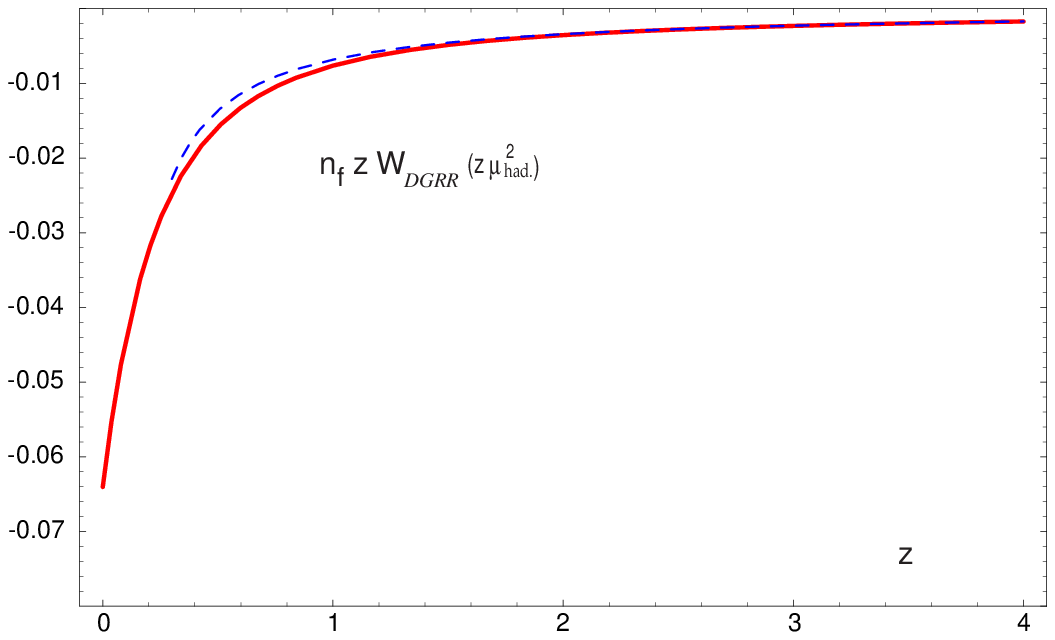}}\vspace*{0.5cm}

\noi{\bf Fig.~5} {\it Shape of
$z\cW_{DGRR}(z\muhad^2)$ (pion pole removed) versus $z$ (solid curve)
for a large--$N_c$ type parameterization as in
Eq.~\rf{largeNDGRR} which matches {\it smoothly} the
short--distance OPE behaviour (the thick dashed line) with the long--distance
$\chi$PT behaviour. The curves correspond to
$\muhad=1~\GeV$, $L_{5}\vert_{\mbox{\tiny\rm large--$N_c$}}=10^{-3}$ and
$\vert\stern\vert^{1/3}=250~\MeV$; the values of the other input parameters
are specified in the Appendix.}

%%%%%%%%%%%% Figure 6 %%%%%%%%%%
\vskip 2pc
\centerline{\epsfbox{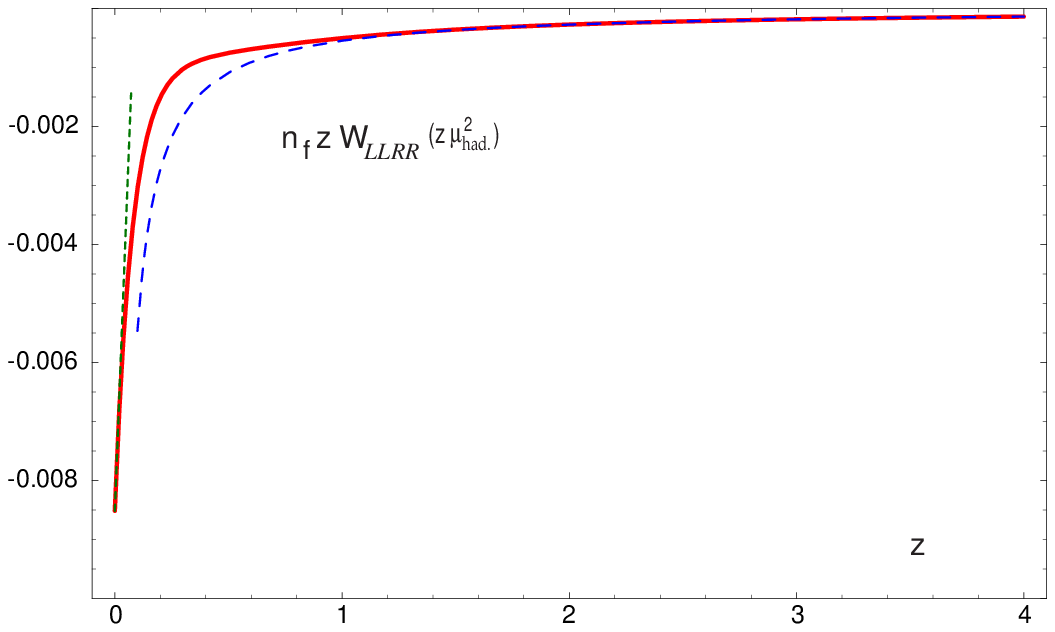}}\vspace*{0.5cm}

\noi
{\bf Fig.~6} {\it Shape of
$z\cW_{LLRR}(z\muhad^2)$ versus $z$ (solid curve) for a large--$N_c$ type
parameterization as in Eq.~\rf{hadfaseLL} which matches
{\it smoothly} the short--distance OPE behaviour (the thick dashed line) with
the long--distance
$\chi$PT behaviour i.e., the point at the origin and the slope (the thin
dashed line). The curves correspond to
$\muhad=1~\GeV$, $L_{5}\vert_{\mbox{\tiny\rm large--$N_c$}}=10^{-3}$ and
$\vert\stern\vert^{1/3}=250~\MeV$; the values of the other input parameters
are specified in the Appendix.}

\vskip 2pc
%%%%%%%%%%%%%%%%%%%%%%%%%%%%%%
\vspace{0.7cm}

%%%%%%%%%%%%%%%%%%%%%%%%%%%%%%%%%%%%%%%%%%%%%%%%%%%%%%%%%
%%%%%%%%%%%%%%%%%%%%%%%%%%%%%%%%%%%%%%%%%%%%%%%%%%%%%%%%%
%%%%%%%%%%%%%%%%%%%%%%%%%%%%%%%%%%%%%%%%%%%%%%%%%%%%%%%%%
\section{\large Analytical results for the coupling
constant $\g_{\underline{8}}$ from QCD Penguins} \setcounter{equation}{0} \label{tcc}
%\section{\large The coupling constant $\g_{\underline{8}}$ from
%QCD Penguins} \setcounter{equation}{0} \lbl{tcc}

\noi
We are now in a position to do the integrals in
Eq.~\rf{g8int} and hence to obtain an evaluation of the
contribution to $\g_{\underline{8}}$ from QCD penguins, beyond the
factorization result in Eq.~\rf{g8fact}. We shall keep the
two--loop evaluation of the Wilson coefficients $C_{4}$ and
$C_{6}$, as obtained e.g. in ref.~\cite{BJLW93}, down to the
charm mass scale $m_c=1.3~\GeV$; while the evolution from $m_c$ to
an arbitrary hadronic scale $\muhad<m_c$ is done at the leading
log approximation, which is consistent with the bosonization of
the $Q_4$ and $Q_6$ operators we have done. This results in the
following expression for $\g_{\underline{8}}\vert_{Q_6,Q_4}$ valid
to leading and next-to-leading order in the $1/N_c$ expansion,
including terms of $\cO(\frac{n_f}{N_c})$:
\be\lbl{g8Q6Q4}
\g_{\underline{8}}\big\vert_{Q_6,Q_4}=C_{6}(m_c)f_{6}(m_c\,; \muhad)+
C_{4}(m_c)f_{4}(m_c\,; \muhad)\,,
\ee
where

{\setl
\bea\lbl{f6}
f_{6}(m_c\,;\muhad) & = & \left(\frac{\als(m_c)}{\als(\muhad)}
\right)^{-\frac{9}{11-2\frac{n_f}{N_c}}}
\left(\frac{\als(m_c)}{\als(\muhad)}
\right)^{\frac{1}{11}\frac{n_f}{N_c}}
\times
\left[-16L_{5}(\muhad)\frac{\stern_{\muhad}^2}{F_{0}^6}\right. \nn \\
& &
\left.
-8n_f\frac{\muhad^4}{16\pi^2F_{0}^4}
\sum_{i}\left(\alpha_{i}\log{\rho}_{i}-\frac{\beta_i}{\rho_{i}}
-\frac{1}{2}\frac{\gamma_{i}}{\rho_{i}^2}\right)
\right]+\frac{1}{9}\frac{n_f}{N_c}\left[1-
\left(\frac{\als(m_c)}{\als(\muhad)}
\right)^{-\frac{9}{11}}\right] \,,
\eea}

\noi
and

{\setl
\bea\lbl{f4}
f_{4}(m_c\,; \muhad) & = &
\left(\frac{\als(m_c)}{\als(\muhad)}
\right)^{\frac{1}{11}\frac{n_f}{N_c}}
\times\left[1+4n_f\frac{\muhad^4}{16\pi^2F_{0}^4}
\sum_{j}\left( \alpha'_{j}\log{\rho}_{j}-\frac{\beta'_j}{\rho_{j}}
-\frac{1}{2}\frac{\gamma'_{j}}{\rho_{j}^{2}}\right)
\right]+
\nn
\\
 & &
\frac{1}{9}\frac{n_f}{N_c}\left[1- \left(\frac{\als(m_c)}{\als(\muhad)}
\right)^{-\frac{9}{11}}\right]\times \left(
-16L_5\frac{\stern^2_{\muhad}}{F_{0}^6} \right)\,.
\eea}

\noi
It is worthwhile to compare the functions $f_{6}(m_c; \muhad)$ and
$f_{4}(m_c; \muhad)$ with the corresponding expressions obtained from
{\it factorization}; i.e.,

{\setl
\bea
f_{6}(m_c\,; \muhad)\big\vert_{\mbox{\rm
{\scriptsize
factorized}}} & = & \left(\frac{\als(m_c)}{\als(\muhad)}
\right)^{-\frac{9}{11}}
\left[-16L_{5}(\muhad)\frac{\stern_{\muhad}^2}{F_{0}^6}\right]
\,,\\
f_{4}(m_c\,; \muhad)\big\vert_{\mbox{\rm
{\scriptsize
factorized}}} & = & 1\,.
\eea}

\noi
We observe that $f_{6}(m_c\,; \muhad)\big\vert_{\mbox{\rm
{\scriptsize
factorized}}}$ has a rather strong dependence on the choice of the
hadronic scale $\muhad$ (a fact which is often ignored in
phenomenological discussions on $\epsilon'/\epsilon$ in the
literature), while $f_{6}(m_c; \muhad)$ has a remarkably smooth
dependence. This behaviour is illustrated in Fig.~7, where we plot
these functions, normalized to their respective values at
$\muhad\!=\!1~\GeV$; i.e.,
\be\lbl{ftilde}
\tilde{f}_{6}(m_c\,; \muhad)\equiv\frac{f_{6}(m_c\,; \muhad)}{f_{6}(m_c;
 1~{\mbox{\rm\footnotesize GeV}})}\quad\annd\quad
\tilde{f}_{6}(m_c; \muhad)_{\mbox{\rm
{\scriptsize
factorized}}}\equiv\frac{f_{6}(m_c;
\muhad)_{\mbox{\rm
{\scriptsize
factorized}}}}{f_{6}(m_c\,;
 1~{\mbox{\rm\footnotesize GeV}})_{\mbox{\rm
{\scriptsize
factorized}}}}
\ee
versus $\muhad$ in the range $0.8~\GeV\le \muhad\le 1.3~\GeV$.
The reason why we normalize the $f_6$ functions to their value at
$\muhad\!=\!1~\GeV$ is that, in absolute value, $f_{6}(m_c\,; \muhad)$
turns out to be  larger than
$f_{6}(m_c; \muhad)_{\mbox{\rm
{\scriptsize
factorized}}}$; for example, at $\muhad\!=0.8~\GeV$, we find that
\be
f_{6}(m_c\,;~0.8~{\mbox{\rm\footnotesize GeV}})\sim 3\times
f_{6}(m_c\,;~0.8~{\mbox{\rm\footnotesize GeV}})_{\mbox{\rm {\scriptsize
factorized}}}\,,
\ee
which is an important enhancement~\footnote{This is in
qualitative agreement with the numerical results found by the authors
of refs.~\cite{BP99,BP00} within the framework of the extended
Nambu-Jona--Lasinio model (ENJL).}.  We also find a similar
enhancement, though perhaps less dramatic, of the unfactorized
contribution to $f_{4}(m_c\,;
\muhad)$, which as we shall see later, is  a welcome feature
towards a phenomenological understanding of the observed $\Delta I=1/2$
rule.

%%%%%%%%%%%% Figure 7 %%%%%%%%%%
\vskip 2pc
\centerline{\epsfbox{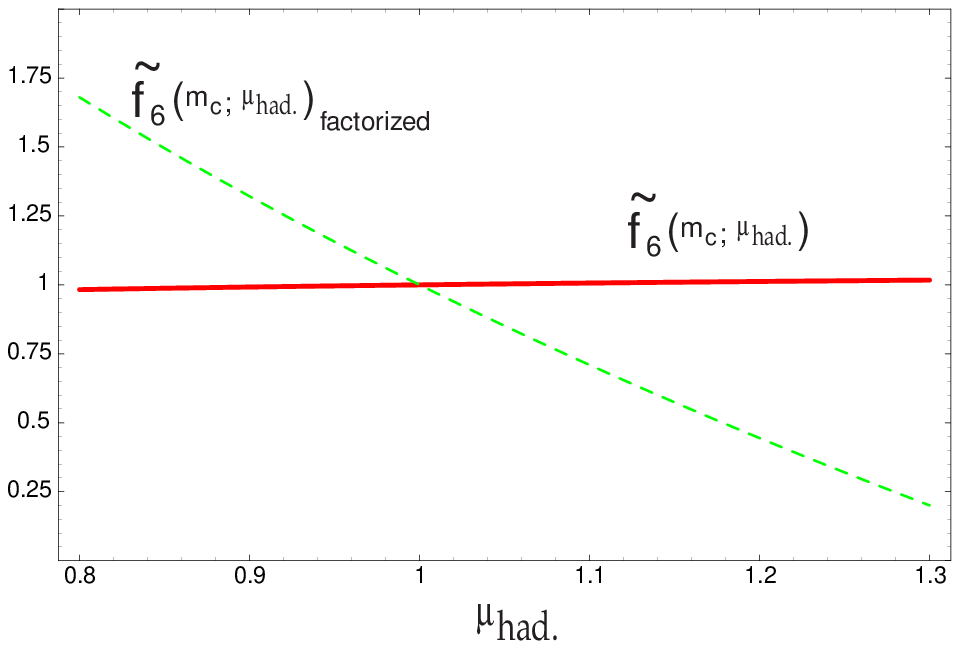}}
%\vspace*{0.5cm}

\noi
{\bf Fig.~7} {\it The dependence on the choice of the $\muhad$ scale
of the functions $f_{6}(m_c;
\muhad)$ (the flat solid curve) and $f_{6}(m_c;
\muhad)_{\mbox{\rm
{\scriptsize
factorized}}}$ (the dotted curve), normalized to their respective values
at
$\muhad\!=\!1~\GeV$. See the definitions in Eq.~\rf{ftilde}.}
%\vskip 2pc
%%%%%%%%%%%%%%%%%%%%%%%%%%%%%%%%

\vspace*{0.3cm}

%%%%%%%%%%%%%%%%%%%%%%%%%%%%%%%%%%%%%%%%%%%%%%%%%%%%%%%%%
%%%%%%%%%%%%%%%%%%%%%%%%%%%%%%%%%%%%%%%%%%%%%%%%%%%%%%%%%
%%%%%%%%%%%%%%%%%%%%%%%%%%%%%%%%%%%%%%%%%%%%%%%%%%%%%%%%%
\section{\large Phenomenology of $K\ra\pi\pi$ Amplitudes and
the $\Delta I=1/2$ Rule}
\setcounter{equation}{0}
\noi
We shall define the decomposition of physical
$K\ra
\pi\pi$ amplitudes  into isospin amplitudes $A_{I=0,2}$, as follows:

{\setl
\bea\lbl{Kpm}
A[K^{0}\ra\pi^{+}\pi^{-}] & = & iA_0
e^{i\delta_0}+\frac{1}{\sqrt{2}}iA_2 e^{i\delta_2}\,, \\
A[K^{0}\ra\pi^{0}\pi^{0}] & = & iA_0
e^{i\delta_0}-\sqrt{2}iA_2 e^{i\delta_2}\,,\lbl{Kzz} \\
A[K^{+}\ra\pi^{+}\pi^{0}] & = & \frac{3}{2}iA_2 e^{i\delta_2}\, ,
\eea}

\noi
where, $\delta_0$ and $\delta_2$ are the $J=0$ $\pi\pi$
phase shifts with $I=0$ and $I=2$ at the $K$ mass. The chiral
Lagrangian in Eq.~\rf{ewcl} gives the following contributions to the
$A_{I}$ isospin amplitudes

{\setl
\bea\lbl{A0}
A_{0} & = & -\frac{\GF}{\sqrt{2}}V_{ud}V_{us}^{*}\ \sqrt{2}F_{0}
\left\{\left(\g_{\underline{8}}+\frac{1}{9}\g_{\underline{27}}
\right)(M_{K}^{2}-m_{\pi}^2)-\frac{2}{3}\frac{1}{F_{0}^4}\
e^2\gew\right\}\,,
\\ A_{2} & = & -\frac{\GF}{\sqrt{2}}V_{ud}V_{us}^{*}\ 2F_{0}\left\{
\frac{5}{9}\g_{\underline{27}}(M_{K}^{2}-m_{\pi}^2)-\frac{1}{3}\frac{1}{F_{0}^4}
\ e^2\gew\right\}\,.
\eea}

\noi
In the large--$N_c$ limit, when only the {\it factorized} contributions
are taken into account, there is a dynamical symmetry which relates the
$\g_{\underline{8}}$ and $\g_{\underline{27}}$ couplings:
\be\lbl{Nc2}
\g_{\underline{8}}\big\vert_{\cO(N_c^2)}=\g_{\underline{27}}
\big\vert_{\cO(N_c^2)}
=\frac{3}{5}\,,
\ee
while we recall that, experimentally,
\be\lbl{expg8g27}
\vert\g_{\underline{8}}\vert_{\mbox{\rm\tiny exp.}}\simeq
5.1\qquad\annd\qquad
\vert\g_{\underline{27}}\vert{\mbox{\rm\tiny exp.}}\simeq 0.29\,:
\ee
although the values to be explained can be
reduced to
\be
\lbl{reducedg8}
\vert\g_{\underline{8}}\vert_{\mbox{\rm\tiny
exp.}}\sim 3.3\quad\annd\quad
\vert\g_{\underline{27}}\vert{\mbox{\rm\tiny exp.}}\sim 0.23\,,
\ee
if one takes into account the enhancement already
provided by the calculated $\cO(p^4)$ chiral
corrections~\cite{KMW91,KDHMW92,PPS01,BDP02}.

The {\it unfactorized} $\cO(N_c)$ contributions break the dynamical
symmetry which is at the origin of the disastrous
prediction in Eq.~\rf{Nc2}; but, as was pointed out in
ref.~\cite{PdeR96}, there still remains a smaller dynamical symmetry at
that level of approximation. This symmetry relates the weak
matrix elements of the
$Q_2$ and $Q_1$ four--quark operators in Eq.~\rf{Q2Q1} (neglecting
their mixing with the penguin operators) to those of the
$\Delta S=2$ operator
\be\lbl{S=2}
Q_{\Delta
S=2}=(\bar{s}_{L}\gamma^{\mu}d_{L})((\bar{s}_{L}\gamma_{\mu}d_{L})\,,
\quad
\langle \bar{K}^{0}\vert Q_{\Delta S=2}
(0)\vert
K^{0}\rangle =f_{K}^2 M_{K}^2\ g_{\Delta S=2}(\mu)\,,
\ee
in the following way:
\be
\Ree\g_{\underline{8}}\big\vert_{Q_2\,,Q_1}=z_{1}(\mu)\left(-1+\frac{3}{5}
g_{\Delta S=2}(\mu)\right)+z_{2}(\mu)\left(1-\frac{2}{5}g_{\Delta
S=2}(\mu)
\right)\,,
\ee
and
\be\lbl{g27}
\g_{\underline{27}}=\left[z_{1}(\mu)+z_{2}(\mu)
\right]\frac{3}{5}g_{\Delta S=2}(\mu)\,,
\ee
where we are using the same definition of the Wilson coefficients as in
Buras {\it et al.}~\cite{BJLW93}; i.e.,
\be\lbl{buras}
C_{i}(\mu)=z_{i}(\mu)+\tau y_{i}(\mu)\,,\quad\with\quad
\tau=-\frac{V_{ts}^{*}V_{td}}{V_{us}^{*}V_{ud}}\,.
\ee
Our present work updates this dynamical symmetry with three new
ingredients:

\begin{enumerate}

\item
The calculation of the constant $g_{\Delta S=2}(\mu)$, within
the same framework of the $1/N_c$ expansion discussed here, which has been
reported in refs.~\cite{PdeR00,CP03}.

\item
Our new result in Eq.~\rf{g8Q6Q4} adds an extra contribution to the
real part of the
$\g_{\underline{8}}$ coupling constant
\be\lbl{eyeQ4Q6}
\Ree
\g_{\underline{8}}\big\vert_{Q_6\,,Q_4}=z_{6}(\mu)f_{6}(\mu\,;\muhad)+
z_{4}(\mu)f_{4}(\mu\,;\muhad)\,,
\ee

\item
The contribution to $\Ree\g_{\underline{8}}$ from the
{\it eye--like} configuration of the $Q_2$ operator, at $\cO(N_c)$, can
be read off straightforwardly from our calculation of the hadronization
of the
$Q_4$ penguin operator, and adds an important extra contribution
\be
\lbl{eyeQ2}
\Ree
\g_{\underline{2}}\big\vert_{Q_2\,,{\mbox{\rm\footnotesize
Eye}}}=z_{2}(\mu)\frac{1}{n_f}\left[ f_{4}(\mu\,;\muhad)-1\right]\,.
\ee
This eye contribution is essential since it provides the $\log{\mu}$
dependence which cancels with the ones from $z_6(\mu) f_6(\mu;\muhad)$ and
$z_4(\mu) f_4(\mu;\muhad)$~\footnote{The $\mu$ scale dependence in
$z_6(\mu)$  and $z_4(\mu)$ induced by the $Q_2$--$Q_{4,6}$ mixing
(multiplied by the factorized piece of $f_{4,6}$ in Eq.~\rf{eyeQ4Q6}) is
cancelled by the $\mu$  scale dependence
of $f_4$ in Eq.~\rf{eyeQ2}.}. Note that this calculation is to our
knowledge the first one
where such a scale dependence cancellation in the $Q_2$--$Q_{4,6}$ mixing
sector is explicitly shown.

\end{enumerate}

\noi
Altogether, and at the $\cO(N_c)$ we are working, we get

{\setl
\bea
\Ree
\g_{\underline{8}} &=& z_{1}(\mu)\left(-1+\frac{3}{5}
g_{\Delta S=2}(\mu)\right)+z_{2}(\mu)\left(1-\frac{2}{5}
g_{\Delta S=2}(\mu)
\right) + z_{2}(\mu) \frac{1}{n_f}\left[
f_{4}(\mu\,;\muhad)-1\right] \nn \\
 & & +z_{6}(\mu)f_{6}(\mu\,;\muhad)+
z_{4}(\mu)f_{4}(\mu\,;\muhad)\,. \lbl{reeg8}
\eea}

The relation in Eq.~\rf{g27} which fixes $\g_{\underline{27}}$ in terms
of $g_{\Delta S=2}(\mu)$ has been known for a long time~\cite{DGH82}
and, in the chiral limit, it holds to all orders in the
$1/N_c$ expansion. In our case, the numerical result for the
invariant $\hat{B}_{K}$ factor~\cite{PdeR00,CP03} obtained in the chiral
limit
\be
\hat{B}_{K}=\frac{3}{4}C_{\Delta S=2}(\mu)\times g_{\Delta
S=2}(\mu)=0.36\pm 0.15
\ee
implies
\be
\g_{\underline{27}}=0.29\pm 0.12\,,
\ee
which is perfectly consistent with the experimental value given in
Eq.~\rf{expg8g27}. This is reassuring, because, at the order of
approximations we are working, it is a complete calculation.
However, for phenomenological applications, it remains to be seen how
this result will be modified in the presence of
chiral corrections.

By contrast, the result for $\Ree \g_{\underline{8}}$ in Eq.~\rf{reeg8}, is not
yet a full $\cO(N_c)$ calculation. We think, however, that it is worthwhile to
present the numerical results which one already obtains at this level. There is no
contribution to $\Ree \g_{\underline{8}}$ from the $Q_{4,6}$ operators for
$\mu\ge m_c$. Therefore, at $\mu=m_c\simeq 1.3~\GeV$, only the first line in
Eq.~\rf{reeg8} is nonvanishing. The corresponding numerical results we obtain for $\Ree
\g_{\underline{8}}\big\vert_{Q_2\,,{\mbox{\rm\footnotesize Eye}}}$, for two input values
of $\vert\stern(\mu=2~\GeV)\vert$ (see discussion below) and letting the {\it
large--$N_c$ value} of $L_5$ vary in the range $1\times 10^{-3}\le
L_{5}\vert_{\mbox{\tiny\rm large--$N_c$}}\le 2\times 10^{-3}$, have been tabulated in
Table~1 below. For each entry, the range of the results is the one corresponding to
Fig.~6 by varying $\muhad$ in the interval $0.8~\GeV\lesssim\muhad\lesssim 1.3~\GeV$,
while, at the same time, allowing for violations of factorization in the residue of the
OPE in Eq.~\rf{opecoeff2} by an extra factor of 2. This extra factor should take into
account the fact that we are finding large deviations from the factorization of matrix
elements of four--quark operators and, therefore, there could be large corrections also
in the residues of the OPE in Eq.~\rf{opecoeff2}.

\begin{table*}[h]
\caption[Results]{Numerical results for  $\Ree
\g_{\underline{8}}\big\vert_{Q_2\,,{\mbox{\rm\footnotesize
Eye}}}$ for various input choices (see text).}
\lbl{table2}
\begin{center}
\begin{tabular}{|c|c|c|c|} \hline \hline
 & & &   \\
$\vert\stern(\mu=2~\GeV)\vert^{1/3} $ &
$L_{5}=1.0\times 10^{-3}$ & $L_{5}=1.5\times 10^{-3}$ & $L_{5}=2.0\times
10^{-3}$  \\
 & & & \\
\hline\hline
$0.260~\GeV$ & 0.35 - 1.23 & 0.44 - 1.21 &  0.59 - 1.17  \\
$0.240~\GeV$ & 0.27 - 1.24 & 0.27 - 1.23 & 0.41 - 1.22  \\
\hline\hline
\end{tabular}
\end{center}
\end{table*}

\noi
From these results, we conclude that a fair estimate of $\Ree
\g_{\underline{8}}$, at the $\cO(N_c)$ we have been working, lies in
the range
\be
\Ree
\g_{\underline{8}}=\underbrace{ 1.33\pm 0.40}_{Q_2,Q_1} +
\underbrace{0.8\pm 0.4}_{Q_{2}-{\mbox{\rm\footnotesize Eye}}}=2.1\pm
0.8\,.
\ee
In spite of the large errors involved, we find this result rather
encouraging. When compared to the value
$\vert\g_{\underline{8}}\vert_{\mbox{\rm\tiny exp.}}\sim 3.3$ to be
explained, it certainly
points in the right direction towards a dynamical understanding of the
$\Delta I=1/2$ enhancement.

%%%%%%%%%%%%%%%%%%%%%%%%%%%%%%%%%%%%%%%%%%%%%%%%%%%%%%%%%
%%%%%%%%%%%%%%%%%%%%%%%%%%%%%%%%%%%%%%%%%%%%%%%%%%%%%%%%%
%%%%%%%%%%%%%%%%%%%%%%%%%%%%%%%%%%%%%%%%%%%%%%%%%%%%%%%%%
%%%%%%%%%%%%%%%%%%%%%%%%%%%%%%%%%%%%%%%%%%%%%%%%%%%%%%%%%
\section{\large Phenomenology of $\epsilon'/\epsilon$}
\setcounter{equation}{0} \lbl{epspheno}

\noi
In terms of
the isospin amplitudes $A_{0}$ and $A_{2}$ in Eqs.~\rf{Kpm} and
\rf{Kzz}, and to a very good approximation, one can write the CP
violation observable
$\varepsilon'/\varepsilon$ as follows, \be
\frac{\varepsilon'}{\varepsilon}=e^{i\Phi}\frac{\omega}
{\sqrt{2}\vert\varepsilon\vert}\left[\frac{\Imm A_{2}}{\Ree A_{2}}
- \frac{\Imm A_{0}}{\Ree A_{0}}\right]\,, \ee with \be \Phi=
\delta_2 -\delta_0 +\frac{\pi}{4}\simeq 0\,,\quad\annd\quad
\omega=\frac{\Ree A_{2}}{\Ree A_{0}}\,. \ee Using now the
effective weak Hamiltonian in Eq.~\rf{effH}, one can obtain a
formal expression of $\varepsilon'/\varepsilon $ in terms of weak
matrix elements of the four--quark operators $Q_{i}$ as
follows~\cite{Buras97}
\be\lbl{mastereppoep}
\frac{\varepsilon'}{\varepsilon}=\Imm\left( V_{ts}^{*}V_{td}
\right)\ \frac{\GF\omega}{2\vert\varepsilon\vert \big\vert\Ree
A_{0}\big\vert}\ \left[ P^{(0)}(1-\Omega_{\mbox{\rm\footnotesize
IB}})-\frac{1}{\omega}P^{(2)} \right]\,,
\ee
where
\be
P^{(I)}=\sum_{i}y_{i}(\mu)\langle (\pi\pi)_{I}\vert
Q_{i}(\mu)\vert K^{0}\rangle\,, \quad\foor\quad I=0,2\,, \ee and
$\Omega_{\mbox{\rm\footnotesize IB}}$ is a term induced by the
effect of isospin breaking $(m_{u}\not= m_{d})$ \be
\Omega_{\mbox{\rm\footnotesize IB}}=\frac{1}{\omega}\frac{(\Imm A_{2})_
{\mbox{\rm\footnotesize IB}}}{\Imm A_{0}}\,.
\ee

It turns out
that, because of the $\Delta I=1/2$ enhancement factor
$1/\omega$, in front of $P^{(2)}$ on the r.h.s. of
Eq.~\rf{mastereppoep}, and because of the values of the Wilson
coefficients $y_{i}(\mu)$, the two $P^{(I)}$ factors  can be
approximated, to a sufficiently good accuracy, as follows
\be
\lbl{P0}
P^{(0)}\simeq y_{6}(\mu)\langle (\pi\pi)_{0}\vert Q_{6}(\mu)\vert
K^{0}\rangle\,+\, y_{4}(\mu)\langle (\pi\pi)_{0}\vert
Q_{4}(\mu)\vert K^{0}\rangle\,, \ee and \be P^{(2)}\simeq
y_{8}(\mu)\langle (\pi\pi)_{2}\vert Q_{8}(\mu)\vert
K^{0}\rangle\,,
\ee
where $Q_{8}$ denotes the electroweak penguin
operator
\be
Q_{8}=-12\sum_{q=u,d,s}e_{q}(\bar{s}_{L}q_{R})(\bar{q}_{R}d_{L})\,,
\ee
with $e_q$ the electric charge of the quark $q$ in $e$ units.

Recall that, since $\tau$ in Eq.~\rf{buras} is complex, the imaginary part of $C_{6}$ and
$C_{4}$, and hence $\Imm \g_{\underline{8}}\vert_{Q_6,Q_4}$ is proportional to $y_6$ and
$y_4$. Our calculation of $g_{\underline{8}}\big\vert_{Q_6,Q_4}$ in section~\ref{tcc},
\normalsize allows, therefore,  for an evaluation of $P^{(0)}$ at the corresponding
approximation i.e., lowest order in $\chi$PT and next--to--leading leading order in the
$1/N_c$ expansion, including terms of $\cO(\frac{n_f}{N_c})$, with the result \be
\lbl{P0f} P^{(0)}\simeq\sqrt{2}F_0 (M_{K}^2-m_{\pi}^2)\left[y_{6}(m_c)f_{6}(m_c\,;
\muhad)+y_{4}(m_c)f_{4}(m_c\,;\muhad) \right]\,, \ee with the long--distance factors
$f_{6}(m_c\,;\muhad)$ and $f_{4}(m_c\,;\muhad)$ given in Eqs.~\rf{f6} and \rf{f4}.

We can also obtain an estimate
of $P^{(2)}$ from our previous work in ref.~\cite{KPdeR01}. As discussed
there, to lowest  $\cO(p^0)$ in the chiral expansion, the
four--quark operator $Q_8$ bosonizes as follows
\be\lbl{Q8bos}
\langle Q_{8}\rangle\vert_{\cO(p^0)} \!=\! -12\hspace*{-0.6cm}
\underbrace{\langle
O_{2}(\mu)\rangle}_{
\langle 0\vert (\bar{s}_{L}s_{R})(\bar{d}_{R}d_{L})\vert
0\rangle}\hspace*{-0.6cm}
\tr\left( U\lambda_{L}^{(23)} U^{\dagger}
 Q_{R}\right)^{\dag}\!\!\,,
\ee where $\lambda_{L}^{(23)}=\delta_{i2}\delta_{j3}$ and $Q_R=
\mbox{\rm diag.}[(2/3,-1/3,-1/3]$. The vev $\langle O_{2}(\mu)\rangle$ also
appears in the Wilson coefficient of the $1/Q^6$ term in the OPE
of the $\Pi_{LR}(Q^2)$ correlation function \be\lbl{LR} \int
d^4x\ e^{iq\cdot x}\langle 0\vert
T\left(\bar{u}_{L}\gamma^{\mu}d_{L}(x)
\bar{u}_{R}\gamma^{\nu}d_{R}(0)^{\dagger}\right)\vert 0\rangle
  = \frac{1}{2i}(q^{\mu}q^{\nu}-g^{\mu\nu}q^2)\Pi_{LR}(Q^2)\,,
\ee
for which the
MHA to large--$N_c$ QCD gives a rather good approximation, as
discussed e.g. in ref.~\cite{deR02}. This offers the possibility
of obtaining an estimate of the vev $O_{2}(\mu)$
\underline{beyond} the strict large--$N_c$ approximation, where
$O_{2}\!\Ra\!\frac{1}{4}\stern^2$, and, therefore, without having
to fix a value for the $\stern$ condensate, which is poorly known
at present. To lowest order in the chiral expansion  ($\cO(p^0)$
in this case as seen from Eq.~\rf{Q8bos}) we find~\cite{Knecht, Peris}
\be
\lbl{P2final}
P^{(2)}\simeq y_{8}(\mu)\left(\frac{8}{3F_{0}^3}
\right)F_{0}^2\frac{M_{V}^2 M_{A}^2}{16\pi\als(\mu)}\times \left[
1-\frac{\als(\mu)}{\pi}\left(\begin{array}{c}25/8
\\  21/8
\end{array} \right) \begin{array}{c}{\mbox{\rm\tiny NDR scheme}}
\\{\mbox{\rm\tiny HV scheme}}
\end{array}\right]\,.
\ee

We have now all the ingredients for a numerical evaluation of $\epsilon'/\epsilon$ in
Eq.~\rf{mastereppoep}. For that purpose, we shall use the values of the physical
parameters given in Appendix A. By far, the most sensitive parameter in our determination
of $\varepsilon'/\varepsilon$ is the QCD quark--condensate. Low values ($\stern^{1/3}\sim
-240~\MeV$) are favoured by various  QCD sum rules determinations, like e.g. the
determination in ref.~\cite{DHGS98}, using $\tau$--data. High values ($\stern^{1/3}\sim
-260~\MeV$) are favoured, however, by some of the lattice QCD simulations~\cite{lattice}.
For two recent determinations see, however, ref.~\cite{new}.  We, therefore,  restrict
the input value of $\stern$ to a range \be\lbl{sternvar}
\vert\stern(\mu=2~\GeV)\vert^{1/3}=(250\pm 10)~\MeV\,.
 \ee Let us recall that the
dependence of $\epsilon'/\epsilon$ on $\stern$ appears in the term $P^{(0)}$, trough the
bosonization of the $Q_6$ operator i.e.,  the function $f_{6}(\mu;\muhad)$ in
Eq.~\rf{f6}, and it is the {\it sixth power} of $\stern^{1/3}$ which counts!

Another
important input parameter is the low--energy constant
$L_5$. As we have already discussed, the $L_5$ which appears in the
factorized contribution induced by the $Q_6$ operator has to be taken
as running, while $L_5$ in the unfactorized contribution is
constant, at the level of accuracy that we are working in the
$1/N_c$--expansion. As we have done in the calculation of $\Ree
\g_{\underline{8}}\big\vert_{Q_2\,,{\mbox{\rm\footnotesize
Eye}}}$ reported above, we leave $L_{5}\vert_{\mbox{\tiny\rm
large--$N_c$}}$ vary in the range $1\times 10^{-3}\le
L_{5}\vert_{\mbox{\tiny\rm large--$N_c$}}\le 2\times 10^{-3}$.

Concerning the factor $\Omega_{\mbox{\rm\footnotesize IB}}$, we have nothing to add at
present. We recall, however, that the estimate has changed from \be
\Omega_{\mbox{\rm\footnotesize IB}}^{\pi^{0}\eta}=0.25\,, \quad {\mbox{\rm See
ref.~\cite{IB}}} \ee to the more recent one $\chi$PT estimate~\cite{EMNP00} \be
\lbl{Omegarange} \Omega_{\mbox{\rm\footnotesize IB}}^{\pi^{0}\eta}=0.16\pm 0.03\,, \ee
which is the value that we shall be using here. We then have that \be\lbl{eepnum}
\frac{\varepsilon'}{\varepsilon}=\underbrace{\Imm\left( V_{ts}^{*}V_{td} \right)\
\frac{\GF\omega}{2\vert\varepsilon\vert \big\vert\Ree A_{0}\big\vert} }_{  (0.055\pm
0.008)\times \mathrm{GeV^{-3}} }\left[ P^{(0)}
\underbrace{(1-\Omega_{\mbox{\rm\footnotesize IB}})}_{0.84\pm
0.03}-\underbrace{\frac{1}{\omega}}_{22.2} P^{(2)} \right]\,. \ee We find large
contributions for both $P^{(0)}$, mostly induced by the $Q_6$ operator (the
contribution from the $Q_4$ operator amounts to less than 3\% of the total), and for
$P^{(2)}$. Typical numerical values are shown in Table~2 below for various input values
of the quark condensate. The values in Table~2 have been obtained by letting
$L_{5}\vert_{\mbox{\tiny\rm large--$N_c$}}$ and $\muhad$ vary in the ranges $1\times
10^{-3}\le L_{5}\vert_{\mbox{\tiny\rm large--$N_c$}}\le 2\times 10^{-3}$  and
$0.8~\GeV\le\muhad\le 1.3~\GeV$. The numbers correspond to the renormalization scale
$\mu=m_c=1.3~\GeV$. The variations induced by the choice $0.8~\GeV\le\mu\le 2~\GeV$
(in $P^{(2)}$), or
$0.7~\GeV\le M_S\le 1.1~\GeV$, change these numbers by $20\%$.

\begin{table*}[h]
\caption[Results]{Numerical results for $P^{(0)} (1-\Omega_{\mbox{\rm\footnotesize IB}})$
in GeV$^3$, $(1/\omega) P^{(2)}$ in GeV$^3$ and $\varepsilon'/ \varepsilon$ for various
input choices of $\stern$. Here $1\times 10^{-3}\le L_{5}\vert_{\mbox{\tiny\rm
large--$N_c$}}\le 2\times 10^{-3}$  and $0.8~\GeV\le\muhad\le 1.3~\GeV$.} \lbl{table1}
\begin{center}
\begin{tabular}{|c|c|c|c|} \hline \hline
 & & &   \\
$\vert\stern(\mu=2~\GeV)\vert^{1/3} $  & $P^{(0)} (1-\Omega_{\mbox{\rm\footnotesize
IB}})\times 10^{2}$ &  $\frac{1}{\omega}
P^{(2)}\times 10^{2}$ &$\frac{\varepsilon'}{\varepsilon}\times 10^{3}$ \\
 & & & \\
\hline\hline
$0.260~\GeV$ & 8.1 - 9.2 & 3.9 &   2.0 - 3.4  \\
$0.250~\GeV$ & 6.3 - 7.2 & 3.9 & 1.2 - 2.1 \\
$0.240~\GeV$ & 5.0 - 5.6 & 3.9 &   0.5 - 1.1  \\
\hline\hline
\end{tabular}
\end{center}
\end{table*}

\noi
The results in Table~2 show that, for the range of values of the quark
condensate $\stern$ in Eq.~\rf{sternvar}, the
values we obtain for $\varepsilon'/\varepsilon$  are perfectly
compatible with  the latest  world average~\cite{WAee} from the NA31,
NA48 and KTeV experiments
\be\lbl{waeep}
\Ree(\epsilon'/\epsilon)\big\vert_{\mbox{\rm\scriptsize Exp.
}}=\left(1.66\pm 0.16\right)\times 10^{-3}\,.
\ee
A vanishing or negative
value
of $\varepsilon'/\varepsilon$, as reported by the lattice QCD
groups~\cite{Soni,Noaki}, appears to us very unlikely.

For the purpose of comparison with other theoretical predictions, it is convenient to
represent Eq.~\rf{eepnum} as a straight line in a plane with $\frac{1}{\omega}P^{(2)}$
plotted in the horizontal axis, and $P^{(0)}(1-\Omega_{\mbox{\rm\footnotesize IB}})$ in
the vertical axis \cite{BG01}. This is the plot shown in Fig.~8, where the width of the
solid line reflects the experimental errors in the overall factor in the r.h.s. of
Eq.~\rf{eepnum} and in Eq.~\rf{waeep}. The cross in the same plot, represents the
prediction which follows from our present estimates of $P^{(0)}$ and
$P^{(2)}$~\footnote{The error in $P^{(2)}$ does not include other analytic
determinations~\cite{Cietal, Bijetal, Narison, Ciri, CDGM02} and lattice QCD
determinations~\cite{Bhatetal,Donetal,Soni, Noaki}.}, for the restricted
input of the quark condensate in Eq.~\rf{sternvar}, but including the
errors of scanning the other input parameters in Eqs.~(A.10) to (A.11) of
the Appendix, and varying the matching scale between $800~\MeV$ and
$2~\GeV$. Furthermore we have also allowed for violations of factorization
in the residue of the OPE in Eq.~\rf{opecoeff1} by a factor of 2, as we
have done before for Eq.~\rf{opecoeff2}. It turns out, however, that the
influence of a possible error in this residue is now much milder because
the UV behaviour of the integral of $Q^2\cW_{DGRR}(Q^2)$ is very much
dominated by the Goldstone double pole term.

\section{\large Discussion and conclusion}
\setcounter{equation}{0}
\lbl{discus}
\noi
We hope to have shown how one can treat analytically, and on a sound
theoretical basis, the calculation of electroweak matrix elements beyond
factorization. In particular, one important goal of this work was to
study the magnitude of the unfactorized contributions in two cases of
special interest, i.e. $\epsilon'/\epsilon$ and the $\Delta I=1/2$ rule.
In section 4, we explained how, bringing new hadronic scales, the
unfactorized contributions may turn out to be sizeable as compared to the
factorized ones, depending on the  scales involved. Already at
the level of the leading ${\cal O}(n_f/N_c)$ corrections, we
have shown by an explicit calculation that this is indeed the case for the
$Q_6$ and $Q_4$ penguin operators.

%%%%%%%%%%%% Figure 8 %%%%%%%%%%
\vskip 2pc
\centerline{\epsfbox{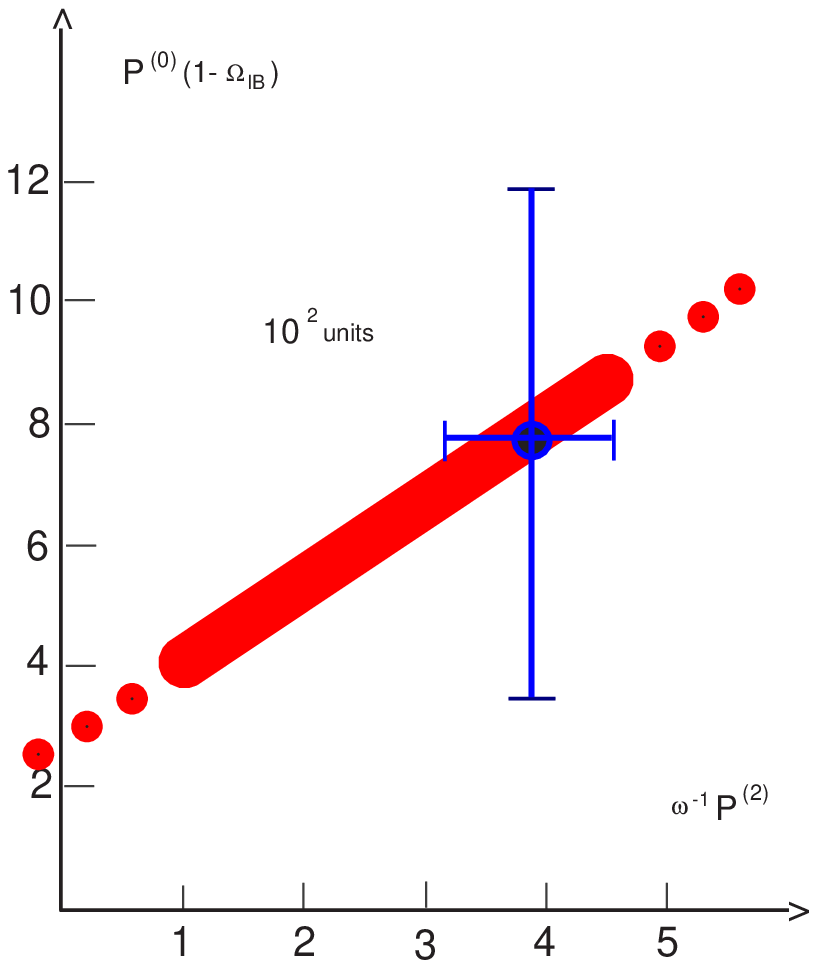}}\vspace*{0.5cm}
{\bf Fig.~8} {\it The experimental value of $\epsilon'/\epsilon$ in
Eq.~\rf{waeep} fixes the straight line in the the figure, with a certain
width due to experimental errors, see Eq.~\rf{eepnum}. Our theoretical
prediction, including errors, is represented by the cross with error
bars.}
\vskip 2pc
%%%%%%%%%%%%%%%%%%%%%%%%%%%%%%

For the operator  $Q_6$, the unfactorized
contribution turns out to be
even much larger than its factorized one, due to the large chiral
coefficient $\sim L_5 - \frac{5}{2} L_3$ in Eq.~\rf{chDG} which
through Eq.~\rf{p4DG} brings a large coefficient to the hadronic scales
in Eq.~\rf{f6}.
As a result, the leading piece is not the factorized contribution
but the unfactorized one for which we just calculated the lowest order
contribution. From this leading lowest order contribution, the higher
order corrections should now be calculated
in subsequent works.
Similarly for $Q_4$ (and for the directly related eye contribution of $Q_2$),
we find a rather large unfactorized contribution through Eqs.~\rf{p2LL} and
\rf{f4}.

Note that the claim that unfactorized contributions are large was
 already made in
refs.~\cite{BBG,HKS99,BP99,PdeR91} for $Q_{1,2}$ and in
refs.~\cite{HKPSB98,HKPS99,BP00}
for $Q_6$. In refs.~\cite{HKPSB98,HKPS99}, the calculation of
the unfactorized
contribution was done for $Q_6$ by just taking the effects of
the pseudoscalar mesons
regularizing the pseudoscalar loops with an euclidean cut-off
$\Lambda_c$. From our
present analysis, which incorporates the effect of hadronic
resonances explicitly, we
observe that this can be justified at the qualitative level,
because, as was anticipated
in those references, a quadratically divergent
( $\sim \Lambda_c^2$) piece can be
interpreted as the reflect of terms proportional to the square of the  masses of the
hadronic resonances. In our more refined approach, the cut--off $\Lambda_c^2$ is replaced
by  combinations of $M_V^2$, $M_S^2$, $M_{P'}^2$ and $\muhad^2$.  This is
also what happens for the quadratic coefficient of the $Q_6$ matrix
element (i.e.~the r.h.s of Eq.~\rf{p4DG} proportional to
$L_5-\frac{5}{2}L_3$).\footnote{Note however that as was already pointed
out above the important $L_3$ contribution was missing in this quadratic
coefficient in ref.~\cite{HKPS99}. As a result the enhancement of the
$Q_6$ contribution to $\varepsilon'/\varepsilon$ found in
ref.~\cite{HKPS99} was smaller than the one we obtain here.} However,
there is no way, with the cut--off procedure advocated in
refs.~\cite{BBG,HKPSB98,HKS99,HKPS99}, to fix unambiguously the choice of
$\Lambda_c$, and there  is no control at all on the short--distance
matching scale dependence. The unfactorized contribution has also been
estimated in refs.~\cite{BF98} within the constituent chiral quark model
of ref.~\cite{PdeR91}, again without short--distance matching. In
refs.~\cite{BP99,BP00}, the previous problem is somewhat ameliorated,
although within the framework of the ENJL model. It turns out, however,
that in this model it is also found that the unfactorized contributions
are large.

The advantage of the large--$N_c$ QCD approach presented here is that it
allows to use the same dimensional regularization both at short and long
distances. We have been able to do a fully analytic matching of scale
dependences, which allows for a clear separation  of the various
scales and hadronic quantum numbers involved in the problem.

To improve our calculation  requires the incorporation of several
effects which are beyond the scope of the present analysis. The ${\cal
O}(1/N_c)$ not enhanced by a $n_f$ factor should be
calculated to have a full NLO calculation in the $1/N_c$ expansion.
The long--distance scheme dependence should also be
calculated to cancel the corresponding short--distance one (which can be
done  exactly in our framework as was shown for $B_K$ in
ref.~\cite{PdeR00}).  The hadronic ansatz could be refined,
systematically, by considering higher dimensional short-- and long--
distance constraints.
Moreover, the effect of the final state interactions (FSI) could
eventually be incorporated, following the work in ref.~\cite{PP00}.
Notice, however, that the effect of the unfactorized contributions which
we find is much larger than the size of the FSI effects estimated in this
reference.~\footnote{In refs.\cite{PP00}, the unfactorized contribution
accounts for an error of
$\pm 0.5 \cdot 10^{-3}$ in their quoted value of
$\varepsilon'/\varepsilon$. We find here that this is an
underestimate.}

Finally, we observe that a precise calculation
of $\varepsilon'/\varepsilon$, within the framework we have discussed
here, is correlated  with the determination of some of the
low energy constants entering our input; in particular the value of
the quark condensate $\stern$ and to a lesser extent the $L_5$ coupling.
Other important input values  are
$\Omega_{IB}$ (where it would be nice to include as well the unfactorized
contribution which, so far,  has been always neglected), $\Lambda_{QCD}$
and Im
$V_{ts}V^*_{td}$.

As a result of all the considerations discussed above, we believe that a systematical
analytic calculation of $\varepsilon'/\varepsilon$ in the Standard Model is now becoming
conceivable although, in our opinion, matching the level of the experimental precision is
likely to become a very difficult task.

\vspace{0.7cm}

{\large{\bf Acknowledgments}}

\vspace{0.3cm}

We thank Matthias Jamin for providing us with his Mathematica code for
the evaluation of the Wilson coefficients, and Hans Bijnens, Marc Knecht,
Laurent Lellouch, Toni Pich and Ximo Prades for helpful discussions. This work
has been supported in part by TMR, EC-Contract No.
HPRN-CT-2002-00311(EURIDICE). The work of S.~Peris has also been partially
supported by the research projects CICYT-FEDER-FPA2002-00748 and
2001-SGR00188. E.~de Rafael is very grateful to ICREA for support during his
stay at the UAB.

%%%%%%%%%%%%%%%%%%%%%%%%%%%%%%%%%%%%%%%%%%%%%%%%%%%%%%%%%
\begin{appendix}
\vspace*{1cm}

\begin{center}
{\bf\large APPENDIX}
\end{center}

\section{\large Compilation of Numerical Inputs}
\setcounter{equation}{0}
\def\theequation{\Alph{section}.\arabic{equation}}

The numerical input we have used in the text are as follows\footnote{Notice that the
$A_{0}$ amplitude in ref.~\cite{Buras97} is {\it ours} times a factor $\sqrt{3/2}$.}:
{\setl \bea
 & &  \Imm\left(
V_{ts}^{*}V_{td}
\right)\
\frac{\GF\omega}{2\vert\varepsilon\vert \big\vert\Ree A_{0}\big\vert}
=0.055\pm0.008\;,\;\mathrm{Ref. \cite{BPS02}.}\\
 & & M_W=80.4~\GeV\,,\quad \sin^2 \theta_{W}=0.23\,,\quad
\alpha=1/137.036\,,
\\ & & m_c=1.3\, \mbox{GeV}\,, \quad m_b=4.4\, \mbox{GeV} \,, \quad m_t=170\,
\mbox{GeV}\,,
\\ & & M_{K}=0.498~\GeV\,,\quad m_{\pi}=0.135~\GeV\,,\quad \omega=
1/22.2\,,\\ & & F_0=0.087\, \mbox{GeV}\,, \quad
M_V=0.770\,\mbox{GeV}\,,
 M_S=0.9 \pm 0.2\,\GeV\,,\\
& & M_{A}=1.2~\GeV\,,\quad M_{V'}=1.4~\GeV\,,\quad M_{P'}=1.3~\GeV\,, \\
& & L_3=-3.0\cdot 10^{-3}\,, \quad L_5(M_\rho)=1.4
\cdot 10^{-3} \,,
\quad L_9(M_\rho) = 6.9  \cdot 10^{-3}\,.
\eea}

and
\begin{eqnarray}
\stern(\mu=2~\GeV) &=& -(250 \pm 10~\MeV)^3\,, \\
L_{5}\vert_{\mbox{\tiny\rm large--$N_c$}}& = & (1.5\pm 0.5)\times
10^{-3}\\
\Lambda_{\msb}^{(4)} &=& (325 \pm 40) \, \mbox{MeV} \,,\\
\Omega_{IB} &=& 0.16 \pm 0.03 \,.
\end{eqnarray}

\noi
The Wilson coefficients we have used have been calculated using the same
program as the one which has been used in ref.~\cite{Buras97}. For
all calculations we used LO Wilson coefficient except for the evaluation of
$P^{(2)}$ in Eq.~\rf{P2final}
for which we took the NLO-NDR Wilson coefficients.
\end{appendix}

%%%%%%%%%%%%%%%%%%%%%%%%%%%%%%%%%%%%%%%%

\end{document}